\documentclass[letterpaper, 10 pt, journal, twoside]{IEEEtran}
\usepackage{cite}
\usepackage{amsmath,amssymb,amsfonts}
\usepackage{array}
\usepackage[caption=false,font=normalsize,labelfont=sf,textfont=sf]{subfig}
\usepackage{graphicx}
\usepackage{textcomp}
\usepackage{stfloats}
\usepackage{url}
\usepackage{hyperref}
\usepackage{verbatim}
\usepackage{hhline}
\usepackage{balance}
\usepackage{soul}
\usepackage{color}
\usepackage{enumitem}
\usepackage{algpseudocode}
\usepackage[ruled,linesnumbered]{algorithm2e}
\usepackage{makecell}
\usepackage{booktabs}

\allowdisplaybreaks

\usepackage[utf8]{inputenc}
\usepackage{multirow}
\bstctlcite{BSTcontrol}

\begin{document}

\title{Integrated Thermal and Power Management for Wave-Powered Subsea Data Centers via Nonlinear Model Predictive Control}

\author{Wanqun Yang and Jun Chen$^{*}$, \IEEEmembership{Senior Member, IEEE}
  \thanks{This work is supported in part by  National Science Foundation through Award \#2430374. \textit{Corresponding author: J. Chen.}}
  \thanks{The authors are with the Department of Electrical and Computer Engineering, Oakland University, Rochester, MI 48309, USA (email: \texttt{\{wanqunyang, junchen\}@oakland.edu}).}%
}

\maketitle

\begin{abstract}
This paper develops an integrated modeling and nonlinear model predictive control (NMPC) framework for coordinating thermal management, flexible workload scheduling, wave-power utilization, and battery operation in a wave-powered subsea data center. Realistic data center workloads are constructed from job-level CPU, memory, and GPU measurements from the MIT Supercloud dataset and divided into interactive and delay-tolerant flexible jobs. 
{Thermal behavior is represented by a three-node lumped model of the IT equipment, recirculating nitrogen, and pressure hull with surrounding seawater as the thermal boundary.} The NMPC jointly optimizes the flexible-workload power budget and cooling command subject to thermal, battery, and workload constraints.
Closed-loop simulations under different workload, thermal, battery, and renewable-generation conditions demonstrate that the proposed framework maintains thermal safety while adapting cooling operation and flexible-workload execution to wave-power availability and battery state-of-charge. The parametric studies show that battery capacity and wave-generation capacity strongly affect battery availability and flexible-workload queue accumulation, while excessive renewable-generation capacity may lead to increased energy curtailment. Monte Carlo and distance-correlation analyses further show that flexible-job delay is relatively insensitive to the investigated system parameters, whereas terminal battery state-of-charge is primarily influenced by battery energy capacity and wave-generation capacity. 

\end{abstract}

{\noindent\textit{Note to Practitioners—}
This work is intended for designers and operators of renewable-powered subsea data centers. The proposed framework coordinates flexible workload execution, cooling operation, wave-power utilization, and battery storage through nonlinear model predictive control. It can be used to evaluate system-level tradeoffs among thermal safety, workload delay, battery sizing, and renewable-generation capacity. The results show that larger battery and wave-generation capacities can improve energy availability and reduce flexible-workload queue accumulation, while excessive renewable capacity may increase energy curtailment. The current study uses simplified thermal and cooling models and a first-come-first-served workload-admission strategy. Therefore, the results are mainly intended to provide system-level design guidance rather than direct hardware operating settings.}


\section{Introduction}
With the rapid development of artificial general intelligence (AGI) in recent years, a growing number of data centers have been constructed or are under development. Moreover, data centers are experiencing rapidly increasing computational demand and energy consumption~\cite{iea2025energyai,mckinsey2025costcompute}. In particular, AGI training requires large scale graphics processing unit (GPU) workloads and high density servers, which significantly increase power demand. Since most of the electrical power consumed by information technology (IT) equipment is eventually converted into heat, the increasing workload intensity also leads to higher thermal management requirements~\cite{chainer2017improving}. Excessive operating temperatures can adversely affect hardware reliability and increase the risk of overheating in IT equipment. As a result, maintaining safe operating temperatures while reducing cooling energy consumption has become an important challenge in next-generation data center design~\cite{garimella2013technological,olivieri2024user}.

Conventional land data centers usually rely on air conditioning, chillers, cooling towers, and air handling units to remove the heat from IT equipment~\cite{moore2006weatherman,meng2020optimization,zhang2021survey}. Although those methods can effectively regulate the temperature, they introduce additional non-IT energy consumption and increase the overall facility power consumption. In some high density data centers, the cooling power consumption can be the main contributor to the energy overhead. For example, \cite{nadjahi2018review} reported that approximately $52\%$ of the electricity is consumed by IT equipment, while about $38\%$ is used by the cooling system and the remaining $10\%$ is consumed by other supporting equipment, such as power distribution and uninterruptible power supply systems. In addition, some cooling technologies, such as evaporative cooling and cooling towers, may also require substantial water consumption, creating additional resource waste and sustainability concerns~\cite{mytton2021data,lei2025water}. Therefore, reducing the cooling overhead and improving resource efficiency have become important directions for next generation data center design.

Microsoft’s Project Natick has demonstrated the feasibility of deploying sealed data center modules underwater and has shown that the subsea environment can provide favorable conditions for data center operation~\cite{cutler2017dunking}. By using the surrounding seawater as a large natural heat sink, subsea data centers can reduce their dependence on conventional chiller-based cooling systems. In the Phase 1 prototype deployment, the system achieved an energy overhead of only approximately 3\%, demonstrating the potential for highly energy-efficient thermal management in subsea data centers~\cite{cutler2017dunking}. However, active cooling is still required to transfer heat from the IT equipment to the external seawater. In a sealed subsea pod, heat is first transferred from the IT equipment to the internal gas environment and is then rejected through gas-to-seawater heat exchangers. This process requires internal circulation fans and seawater pumps, while additional passive heat transfer occurs through the pressure hull. Therefore, the cooling system must be controlled to maintain thermal safety without introducing unnecessary fan and pump power consumption.

The power supply of a subsea data center presents an additional operational challenge. Reliance on an onshore power grid requires subsea power transmission and dedicated electrical infrastructure, increasing the complexity of offshore power delivery~\cite{microsoft_natick_phase2,green2007electrical,rajashekara2017electrification}. Local marine renewable generation can reduce this dependence, and wave energy is particularly suitable because it is available offshore and can be deployed close to the subsea data center~\cite{falcao2010wave}. However, wave-power generation varies over time and does not necessarily coincide with the data center load. Battery storage can buffer short-term power imbalances, but its available energy is limited by the battery state-of-charge (SOC)~\cite{kaheni2024rule}. Consequently, the data center workload, cooling demand, renewable generation, and battery operation must be coordinated rather than considered independently.

Data center workloads contain both time-sensitive and delay-tolerant jobs~\cite{rostami2024linearized}.Interactive workloads must be executed when they arrive, whereas flexible workloads may be postponed, enabling energy-aware scheduling of deferrable computational tasks~\cite{wang2025deep,lin2023autoencoder}. This workload flexibility provides an opportunity to adapt the IT power demand to the available renewable energy and battery condition~\cite{han2023two,zhang2024recursive}. Nevertheless, executing more flexible workload increases both the electrical demand and the heat generated by the IT equipment. Increasing the cooling command improves heat rejection but also increases fan and pump power consumption. Therefore, workload scheduling, cooling operation, battery SOC and thermal dynamics are strongly coupled~\cite{rostami2024linearized}. Such a coupled system requires an effective control method to coordinate different operating objectives and constraints. Model predictive control (MPC) has been widely used because it can predict future system behavior and determine control actions through optimization~\cite{li2025hierarchical}, and it has also been applied to data center energy and thermal management~\cite{ogawa2015development,ogura2018model}. For systems with nonlinear dynamics, nonlinear model predictive control (NMPC) can directly incorporate the nonlinear system model into the optimization problem. 

{Recent studies have addressed individual aspects of this coupled problem. Rostami et al.~\cite{rostami2024linearized} considered coordinated data-center workload and cooling management, while Han et al.~\cite{han2023two} investigated demand response for flexible data-center operation. Predictive battery scheduling under renewable-generation and load uncertainties has also been studied in microgrids~\cite{kaheni2024rule}. However, these studies do not simultaneously coordinate job-level flexible workload execution, active cooling, renewable generation, and battery storage. This gap motivates the integrated NMPC framework developed here for wave-powered subsea data centers.}

A cooling-only controller does not exploit the flexibility of delay-tolerant jobs, while workload scheduling without thermal and energy awareness may cause excessive battery depletion, high cooling demand, or thermal-constraint violations. To address these coupled requirements, this paper develops an integrated modeling and control framework for a wave-powered subsea data center. Job-level CPU, memory, and GPU operational data from the MIT Supercloud dataset are used to construct interactive- and flexible-workload power profiles. A three-node thermal model describes the heat transfer among the IT equipment, recirculating nitrogen, pressure hull, and surrounding seawater. Based on the coupled workload, thermal, cooling, wave-power, and battery models, an NMPC controller jointly determines the flexible-workload power budget and normalized cooling command. The optimized flexible-workload budget is then applied to the waiting jobs using a strict first-come-first-served (FCFS) admission rule. The waiting jobs are ordered according to their arrival times and considered sequentially from the head of the queue. A job is admitted only when its first-step power requirement can be accommodated within the flexible-workload power budget and the battery-energy reserve condition is satisfied.

The main contributions of this paper are summarized as follows:

\begin{enumerate}

\item A workload-driven subsea data center model is developed using job-level CPU, memory, and GPU data from the MIT Supercloud dataset. The workload is divided into interactive and delay-tolerant flexible jobs, allowing the actual job-level workload characteristics to be incorporated into the thermal and energy management framework.
\item An integrated thermal-energy model is developed to describe the coupled operation of the IT equipment, recirculating nitrogen, pressure hull, active cooling system, wave generation, battery storage, and flexible-workload queue.
\item An NMPC-based control framework is developed to coordinate flexible-workload execution and active cooling by jointly optimizing the flexible-workload power budget and cooling command. The optimized workload budget is further implemented through a strict FCFS job-admission rule that connects the continuous NMPC decision with discrete job-level execution.
\item The proposed framework is evaluated under different workload, thermal, battery, and wave-generation conditions. Parametric studies together with Monte Carlo simulation and distance-correlation analyses are conducted to investigate how these factors affect thermal performance, flexible-job scheduling, and battery operation, and to identify the system parameters that have stronger effects on the closed-loop performance.

\end{enumerate}

The rest of this paper is organized as follows. Section~\ref{sec-system} presents the overall system configuration. Section~\ref{secsub-IT_workload} develops the subsea data center workload model. Section IV presents the subsea data center thermal model. Section V describes the cooling-regulation and power-consumption models. Section~\ref{sec-energy_modeling} presents the energy-supply model, including wave-power generation and battery storage. Section~\ref{sec-optimization} formulates the NMPC-based cooling and workload control framework. Section~\ref{sec-results} presents the simulation results and parametric analyses. Section~\ref{sec-conclusion} concludes the paper.

\section{System Overview}\label{sec-system}

\begin{figure*}[ptb]
    \centering
    \includegraphics[width=0.7\linewidth]{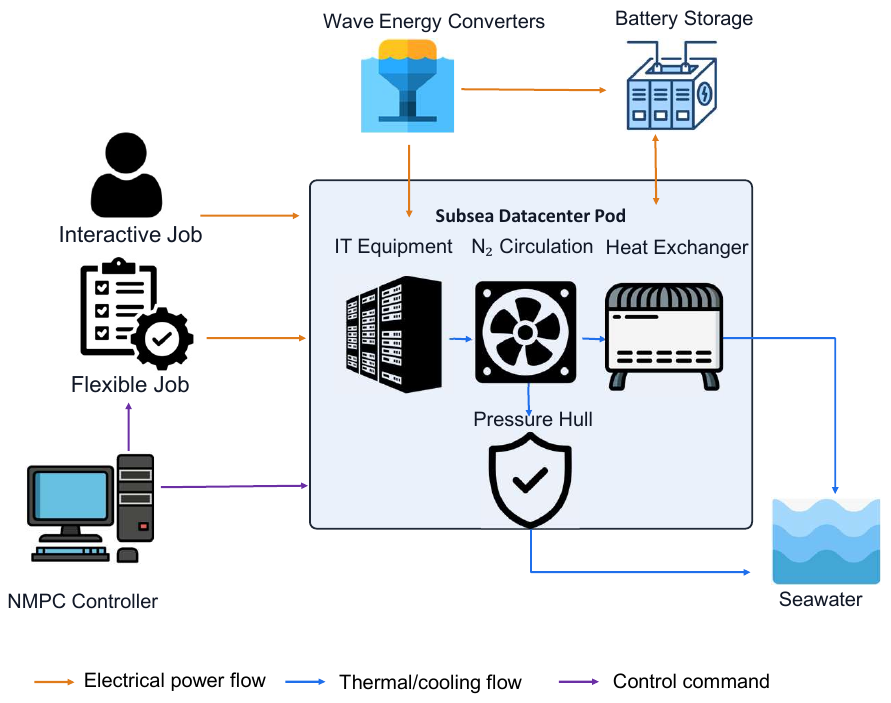}
    \caption{System configuration of the subsea data center under study.}
    \label{fig:system_configuration}
\end{figure*}

Figure~\ref{fig:system_configuration} illustrates the overall configuration of the proposed wave-powered subsea data center system. The data center workload is constructed from the MIT Supercloud dataset and is divided into interactive and flexible jobs~\cite{samsi2021supercloud}. Interactive jobs are executed upon arrival, whereas flexible jobs can be deferred and scheduled by the NMPC controller. The controller then coordinates the flexible-job execution and the cooling operation of the subsea data center pod for performance optimization. 

On the power side, wave energy converters provide renewable electricity to the data center and battery storage system. The battery stores surplus wave energy and supplies power when the wave-power generation is insufficient to meet the data center demand. Inside the hull, the IT equipment converts the supplied electrical power into heat. The generated heat is transferred to the recirculating nitrogen and then removed through the heat exchanger to the surrounding seawater. Heat can also be passively transferred through the pressure hull and rejected to the seawater. In Fig.~\ref{fig:system_configuration}, the orange, blue, and purple arrows represent electrical-power flow, thermal/cooling flow, and control commands, respectively. 

\section{Subsea Data Center Workload Model}\label{secsub-IT_workload}
A data-driven workload model is constructed from the operational logs and hardware-monitoring records of the MIT Supercloud high-performance computing cluster~\cite{samsi2021supercloud}. Since the electrical power consumed by the IT equipment is ultimately converted into heat, the IT power demand is treated as the primary heat source in the thermal model~\cite{chainer2017improving}. To construct a time-aligned IT power profile, the framework processes job-level CPU utilization, memory usage, and GPU power, while the Slurm scheduler logs provide the execution intervals and allocated resources of individual jobs.
The total transient IT power supplied to the thermal model is then expressed as
\begin{equation}
\label{eq:P_it}
P_{\mathrm{IT}}(t)=P_{\mathrm{base}}+P_{\mathrm{job}}(t),
\end{equation}
where $P_{\mathrm{base}}$ represents the background IT power that is not explicitly captured by the job-level workload data, including idle-server, networking, and other supporting electronic loads.
The workload-dependent power is defined as
\begin{equation}
\label{eq:P_job}
P_{\mathrm{job}}(t)=P_{\mathrm{CPU}}(t)+P_{\mathrm{MEM}}(t)+P_{\mathrm{GPU}}(t),
\end{equation}
where $P_{\mathrm{CPU}}(t)$, $P_{\mathrm{MEM}}(t)$, and $P_{\mathrm{GPU}}(t)$ denote the CPU power, memory power, and measured GPU power associated with the active jobs, respectively.

The workload-dependent CPU power $P_{\mathrm{CPU}}$ is estimated using a utilization-scaled per-core power model, following the general methodology adopted in computation-energy estimation
studies~\cite{lannelongue2021green}. The MIT Supercloud employs Intel Xeon Gold 6248 and Xeon Platinum 8260 processors. According to the manufacturer specifications, the Xeon Gold 6248 has 20 physical cores and a TDP of 150 W \cite{intel_xeon_gold_6248}, whereas the Xeon Platinum 8260 has 24 physical cores and a TDP of 165 W \cite{intel_xeon_platinum_8260}. Dividing the nominal TDP values by the corresponding physical core counts gives approximately 7.50 W/core and 6.88 W/core, respectively. These TDP-based values include package-level static, cache, and uncore power. In the present model, idle-server and other background IT consumption are represented separately through $P_{\mathrm{base}}$. Therefore, the workload-dependent CPU power coefficient is conservatively set to $k_{\mathrm{CPU}}=5~\mathrm{W/core}$. 
The aggregate workload-dependent CPU power is calculated as
\begin{equation}
\label{eq:P_cpu}
P_{\mathrm{CPU}}(t)
=
k_{\mathrm{CPU}}N_{\mathrm{core}}(t),
\end{equation}
where $N_{\mathrm{core}}(t)$ denotes the total number of CPU cores utilized by all active jobs at time $t$. It is derived from the job-level CPU utilization records, in which a utilization value of 100\% corresponds to one fully utilized CPU core. For each job, the estimated number of utilized cores is limited by the number of CPU cores allocated to that job according to the Slurm scheduler records. 

Memory power is estimated from the total resident memory usage of all
active jobs:
\begin{equation}
\label{eq:P_mem}
P_{\mathrm{MEM}}(t)=k_{\mathrm{mem}} M_{\mathrm{total}}(t),
\end{equation}
where $M_{\mathrm{total}}(t)$ denotes the total resident set size (RSS) of all jobs active at time $t$, expressed in GB. The RSS measurements are obtained from the job-level monitoring records and aligned with the corresponding job execution intervals using the timestamps provided by the Slurm scheduler logs. Following the linear memory-power approximation adopted in \cite{budennyy2022eco2ai}, in which the power associated with process-allocated memory is estimated using a coefficient of $0.375~\mathrm{W/GB}$, the workload-dependent memory power coefficient is approximated as $k_{\mathrm{mem}}=0.4~\mathrm{W/GB}$ in this work. 

Unlike CPU and memory power, GPU power is obtained directly from hardware telemetry logs rather than estimated from resource utilization. GPU power profiles are aggregated across all GPUs associated with active jobs to obtain the total GPU power $P_{\mathrm{GPU}}(t)$. The CPU power, memory power, and measured GPU power are then combined to form the workload-dependent power $P_{\mathrm{job}}(t)$, as defined in~\eqref{eq:P_job}. The total IT power $P_{\mathrm{IT}}(t)$ is subsequently obtained by adding the background IT power $P_{\mathrm{base}}$, as given in~\eqref{eq:P_it}. 

For subsequent workload-scheduling and closed-loop simulations, the processed data are stored. The saved data include each job’s execution time, job type, allocated resources, power profile relative to its job start, power profile aligned with its original execution time, CPU, memory, GPU, and total energy consumption, together with the corresponding dataset metadata. Table~\ref{tab:thermal_params} lists all parameter values and their sources used in thermal model.


\section{Subsea Data Center Thermal Model}

\begin{table}[ptb]
\centering
\caption{Thermal-model parameters.}
\label{tab:thermal_params}
\begin{tabular}{lll}
\hline
Parameter & Value & Source \\
\hline
\multicolumn{3}{l}{\textit{IT node}}\\
$N_{\mathrm{rack}}$              & 44                        & \cite{samsi2021supercloud} \\
$m_{\mathrm{rack}}$              & $900~\mathrm{kg}$         &System Design \\
$c_{\mathrm{IT}}$                & $600~\mathrm{J/(kg\,K)}$  & \cite{ibrahim2012thermal} \\
$\varepsilon_{\mathrm{IT}}$      & $0.90$                    & \cite{pardey2014further} \\
\hline
\multicolumn{3}{l}{\textit{Nitrogen node}}\\
$c_{\mathrm{N_2}}$               & $1040~\mathrm{J/(kg\,K)}$ &  \cite{lin2016pulsed} \\
$\rho_{\mathrm{N_2}}$            & $1.165~\mathrm{kg/m^3}$   & \cite{lin2016pulsed} \\
$D_{\mathrm{pod}}$               & $4.0~\mathrm{m}$          &System Design \\
$L_{\mathrm{pod}}$               & $45.0~\mathrm{m}$         & System Design \\
$A_{\mathrm{h}}$                 & $590.6~\mathrm{m^2}$      & Derived From \eqref{eq:hull_area}\\
$V_{\mathrm{cab}}$               & $565.5~\mathrm{m^3}$      & Derived From \eqref{eq:V_cab}\\
$m_{\mathrm{N_2}}$               & $658.6~\mathrm{kg}$       & 
 Derived From \eqref{eq:m_n2}\\
$h_{\mathrm{N_2,h}}$             & $4.0~\mathrm{W/(m^2\,K)}$ & \cite{moran2024modeling} \\
\hline
\multicolumn{3}{l}{\textit{Pressure-hull node}}\\
$t_{\mathrm{h}}$                 & $0.04~\mathrm{m}$         & System Design \\
$\rho_{\mathrm{h}}$              & $7850~\mathrm{kg/m^3}$    & \cite{feng2020efficient} \\
$c_{\mathrm{h}}$                 & $470~\mathrm{J/(kg\,K)}$  & \cite{gao2024oil} \\
$h_{\mathrm{h,sw}}$              & $400~\mathrm{W/(m^2\,K)}$ & \cite{santinello2017external} \\
\hline
\multicolumn{3}{l}{\textit{Seawater / heat exchanger}}\\
$c_{\mathrm{sw}}$                & $3991.87~\mathrm{J/(kg\,K)}$ &  \cite{kondeti2025estimating}\\
$\varepsilon_{\mathrm{HX}}$      & $0.425$--$0.690$ (map)    & \cite{schmidt2005maintaining,gao2014dynamic} \\
HX $\varepsilon$-model           & --                        & \cite{navarro2007effectiveness} \\
\hline
\end{tabular}
\end{table}

To describe the thermal dynamics shown in Fig.~\ref{fig:system_configuration}, a three-node lumped-parameter thermal model is developed. The three thermal states are the IT equipment temperature $T_{\mathrm{IT}}$, the recirculating nitrogen temperature $T_{\mathrm{N_2}}$, and the pressure hull temperature $T_{\mathrm{h}}$. The seawater temperature $T_{\mathrm{sw}}$ is treated as a time-varying boundary condition. The heat generated by the IT equipment is first transferred to the recirculating nitrogen. Part of the heat carried by the nitrogen is removed by the gas-to-seawater heat exchanger. The remaining heat can be transferred to the pressure hull. 
\subsection{IT Temperature (${T}_{\mathrm{IT}}$)}
The energy stored in the IT node changes according to the IT heat-generation rate and the heat exchanged between the IT equipment and the recirculating nitrogen. The corresponding governing equation is
\begin{equation}
C_{\mathrm{IT}}\dot{T}_{\mathrm{IT}}=
P_{\mathrm{IT}}-\dot{Q}_{\mathrm{IT,N_2}},
\label{eq:IT_balance}
\end{equation}
where $C_{\mathrm{IT}}$ is the thermal capacitance, calculated as
\begin{equation} 
C_{\mathrm{IT}} = N_{\mathrm{rack}} m_{\mathrm{rack}} c_{\mathrm{IT}}, \label{eq:C_IT} 
\end{equation} 
where $N_{\mathrm{rack}}$ is the number of racks. $m_{\mathrm{rack}}$ is the effective thermal mass of one rack. Since the detailed hardware composition of each rack is unavailable, $m_{\mathrm{rack}}$ is treated as an assumed effective value. The equivalent specific heat capacity $ c_{\mathrm{IT}}$ is set to \(600~\mathrm{J/(kg\,K)}\). 
$\dot{Q}_{\mathrm{IT,N_2}}$ is the heat-transfer rate between IT Equipment and Nitrogen flow, defined as 
\begin{equation} 
\dot{Q}_{\mathrm{IT,N_2}} = G_{\mathrm{IT}} \left( T_{\mathrm{IT}} - T_{\mathrm{N_2}} \right). 
\label{eq:Q_IT_N2} 
\end{equation} 
The effective thermal conductance $G_{\mathrm{IT}}$ depends on the total nitrogen mass-flow rate and is calculated as 
\begin{equation}
G_{\mathrm{IT}} = \varepsilon_{\mathrm{IT}} \dot{m}_{\mathrm{N_2}} c_{\mathrm{N_2}}, 
\label{eq:G_IT} 
\end{equation} 
where $\dot{m}_{\mathrm{N_2}}$ is the total nitrogen mass-flow rate, $c_{\mathrm{N_2}}$ is the specific heat capacity of nitrogen, and $\varepsilon_{\mathrm{IT}}$ is an effective heat-transfer factor. 

\subsection{Nitrogen Temperature (${T}_{\mathrm{N_2}}$)}
The nitrogen node exchanges heat with the IT equipment, the gas-to-seawater heat exchangers, and the pressure hull. The nitrogen energy balance is
\begin{equation}
C_{\mathrm{N_2}}\dot{T}_{\mathrm{N_2}}=\dot{Q}_{\mathrm{IT,N_2}}-\dot{Q}_{\mathrm{HX}}-\dot{Q}_{\mathrm{N_2,h}}.
\label{eq:N2_balance}
\end{equation}
The thermal capacitance of the recirculating nitrogen $C_{\mathrm{N_2}}$ is calculated as  
\begin{equation} 
C_{\mathrm{N_2}} = m_{\mathrm{N_2}} c_{\mathrm{N_2}},
\label{eq:C_N2} 
\end{equation}
where $c_{\mathrm{N_2}}$ is the specific heat capacity of nitrogen at constant pressure and $m_{\mathrm{N_2}}$ is total nitrogen mass in the subsea data center pod calculated by 
\begin{equation}\label{eq:m_n2}
    m_{\mathrm{N_2}} = \rho_{\mathrm{N_2}} V_{\mathrm{cab}}.
\end{equation}
Here $\rho_{\mathrm{N_2}}$ is nitrogen density. Since the subsea data center pod is approximated as a cylindrical enclosure, its internal volume is 
\begin{equation}\label{eq:V_cab}
    V_{\mathrm{cab}} = \pi D_{\mathrm{pod}}^{2} L_{\mathrm{pod}}/4,
\end{equation}
where $D_{\mathrm{pod}}$ and $L_{\mathrm{pod}}$ are the pod diameter and length, respectively. 
Let $\dot{Q}_{\mathrm{HX}}$, where subscript HX denotes the ``heat exchanger'', is heat transfer rate of recirculating nitrogen and the seawater and is defined as 
\begin{equation} 
\dot{Q}_{\mathrm{HX}} = G_{\mathrm{HX}} \left( T_{\mathrm{N_2}} - T_{\mathrm{sw}} \right), 
\label{eq:Q_HX} 
\end{equation}
where heat-exchanger thermal conductance $G_{\mathrm{HX}}$ is calculated by
\begin{equation}
G_{\mathrm{HX}}=
N_{\mathrm{HX,act}}
\varepsilon_{\mathrm{HX}}
C_{\min},
\label{eq:G_HX}
\end{equation}
where \(N_{\mathrm{HX,act}}\) is the equivalent number of active HX units. In the subsequent control optimization, only a fraction of the HX capacity may be required at a given operating condition. Therefore, \(N_{\mathrm{HX,act}}\) is treated as a continuous equivalent quantity and may take non-integer values rather than representing the actual number of operating HX units. In the effectiveness-based HX model~\cite{navarro2007effectiveness}, the heat transfer capacity is limited by the smaller heat capacity rate $C_{\min}$ between
the nitrogen side and the seawater side. Therefore, to obtain total equivalent HX conductance $G_{\mathrm{HX}}$, the heat capacity rate of each side is first calculated. First, the nitrogen side heat capacity rate is defined as 
\begin{equation} 
C_{\mathrm{g}} = \dot{m}_{\mathrm{N_2,u}} c_{\mathrm{N_2}}, 
\label{eq:gas_capacity_rate} 
\end{equation}
where \(\dot{m}_{\mathrm{N_2,u}}\) is the nitrogen mass-flow rate through one active heat-exchanger unit. It is related to the total recirculating nitrogen mass-flow rate by $\dot{m}_{\mathrm{N_2,u}} = \dot{m}_{\mathrm{N_2}} / N_{\mathrm{HX,act}}$. 
The seawater-side heat-capacity rate is calculated as 
\begin{equation} C_{\mathrm{sw}} = \dot{m}_{\mathrm{sw,u}} c_{\mathrm{sw}}, 
\label{eq:sea_capacity_rate} 
\end{equation} 
where \(\dot{m}_{\mathrm{sw,u}}\) is the seawater mass-flow rate through one active heat-exchanger. Similarly, $\dot{m}_{\mathrm{sw,u}} = \dot{m}_{\mathrm{sw}} / N_{\mathrm{HX,act}}$, where  $\dot{m}_{\mathrm{sw}}$ is the total seawater flow rate.  $c_{\mathrm{sw}}$ is the specific heat capacity of seawater. The smaller heat capacity rate is then defined as 
\begin{equation} 
C_{\min} = \min \left( C_{\mathrm{g}}, C_{\mathrm{sw}} \right). 
\label{eq:C_min} 
\end{equation} 
The heat-exchanger effectiveness $\varepsilon_{\mathrm{HX}}$ in ~\eqref{eq:G_HX} is taken from the IBM rear-door heat-exchanger data of~\cite{schmidt2005maintaining}, which also had been used in data center thermal study~\cite{gao2014dynamic}. The air-to-water effectiveness map in \cite{schmidt2005maintaining} is used as a surrogate for the nitrogen-to-seawater HX, which is justified because nitrogen and air have similar thermophysical properties, as do seawater and fresh water. For each operating point, $\varepsilon_{\mathrm{HX}}$ is obtained by two-dimensional linear interpolation of this map against the per-unit nitrogen and seawater mass flow rates. 

The last term $\dot{Q}_{\mathrm{N_2,h}}$ in Nitrogen temperature dynamic~\eqref{eq:N2_balance} is nitrogen exchanges heat with the pressure hull, it is defined as 
\begin{equation} 
\dot{Q}_{\mathrm{N_2,h}} = G_{\mathrm{N_2,h}} \left( T_{\mathrm{N_2}} - T_{\mathrm{h}} \right).
\label{eq:Q_N2_h} 
\end{equation}
The equivalent thermal conductance between the nitrogen and the pressure hull is calculated as 
\begin{equation} 
G_{\mathrm{N_2,h}} = h_{\mathrm{N_2,h}} A_{\mathrm{h}}, 
\label{eq:G_N2_h} 
\end{equation} 
where \(h_{\mathrm{N_2,h}}\) is the effective heat-transfer coefficient between the nitrogen and the internal surface of the pressure hull. \cite{moran2024modeling} investigated heat transfer in a commercial cryostat containing nitrogen gas between stainless-steel surfaces, and the nitrogen-side convective heat-transfer coefficients shown in their figure are approximately 0.6--8$~\mathrm{W/(m^2K)}$. \(A_{\mathrm{h}}\) is the hull surface area. The pressure hull is approximated as a closed cylindrical enclosure.
Its total surface area, including the lateral surface and two end
surfaces, is calculated as
\begin{equation}
A_{\mathrm{h}}
=
\pi D_{\mathrm{pod}}L_{\mathrm{pod}}
+
\frac{\pi D_{\mathrm{pod}}^{2}}{2}.
\label{eq:hull_area}
\end{equation}

\subsection{Pressure Hull Temperature (${T}_{\mathrm{h}}$)}
The pressure hull exchanges heat with the recirculating nitrogen and the surrounding seawater. The temperature dynamic of hull is
\begin{equation}
C_{\mathrm{h}}\dot{T}_{\mathrm{h}} = \dot{Q}_{\mathrm{N_2,h}}-\dot{Q}_{\mathrm{h,sw}}
\label{eq:hull_balance}
\end{equation}
The thermal capacitance of the pressure hull is expressed as 
\begin{equation} 
C_{\mathrm{h}} = A_{\mathrm{h}} t_{\mathrm{h}} \rho_{\mathrm{h}} c_{\mathrm{h}}, 
\label{eq:C_hull} 
\end{equation} 
where $t_{\mathrm{h}}$ is the structural thickness, $\rho_{\mathrm{h}}$ is the hull material density, and $c_{\mathrm{h}}$ is the specific heat capacity of the hull material. 
The pressure hull also exchanges heat with the surrounding seawater. The corresponding heat-transfer rate is defined as 
\begin{equation} 
\dot{Q}_{\mathrm{h,sw}} = G_{\mathrm{h,sw}} \left( T_{\mathrm{h}} - T_{\mathrm{sw}} \right). 
\label{eq:Q_h_sw} 
\end{equation}
where 
    $G_{\mathrm{h,sw}}=h_{\mathrm{h,sw}}A_{\mathrm{h}}$
 is the effective thermal conductance between the pressure hull and the surrounding seawater, and $h_{\mathrm{h,sw}}$ is the hull-to-seawater heat-transfer coefficient. \cite{santinello2017external} studied heat transfer from a submerged cylindrical hull to seawater using empirical correlations and CFD simulations, obtaining heat-transfer coefficients of approximately 360--670.5$~\mathrm{W/(m^2K)}$. $A_{\mathrm{h}}$ is the hull surface area defined in \eqref{eq:hull_area}. 

\section{Data center Cooling Regulation and Power Consumption}
\subsection{Cooling Flow Regulation}\label{secsub-cooling_control}
\begin{table}[t]
\centering
\caption{Cooling-regulation and power-model parameters.}
\label{tab:cooling_params}
\begin{tabular}{lll}
\hline
Parameter & Value & Source \\
\hline
\multicolumn{3}{l}{\textit{Flow regulation}}\\
$f_{\min}$                       & $1/12 \approx 0.083$          & System Design \\
$N_{\mathrm{HX}}$                & 44                            & System Design \\
$N_{\mathrm{fan}}$               & 2688                          & System Design \\
$n_{\mathrm{rated}}$    & $13000~\mathrm{rpm}$          & \cite{delta_ffb0412shn} \\
$\dot{V}_{\mathrm{fan,rated}}$   & $0.680~\mathrm{m^3/min}$      & \cite{delta_ffb0412shn} \\
$\dot{m}_{\mathrm{sw,u,min}}$    & $0.441~\mathrm{kg/s}$         & \cite{schmidt2005maintaining}  \\
$\dot{m}_{\mathrm{sw,u,max}}$    & $0.693~\mathrm{kg/s}$         & \cite{schmidt2005maintaining} \\
\hline
\multicolumn{3}{l}{\textit{Fan power}}\\
$a_1$                            & $2.46\times10^{-4}$           & \cite{garraghan2016unified} \\
$a_2$                            & $-3.70\times10^{-8}$          & \cite{garraghan2016unified} \\
$a_3$                            & $3.11\times10^{-12}$          & \cite{garraghan2016unified} \\
\hline
\multicolumn{3}{l}{\textit{Seawater pump}}\\
$N_{\mathrm{pump}}$              & 4                             & System Design\\
$Q_{\mathrm{pump,ref}}$          & $27.0~\mathrm{m^3/h}$         & \cite{calpeda_nms_catalog} \\
$P_{\mathrm{pump,ref}}$          & $2.2~\mathrm{kW}$             & \cite{calpeda_nms_catalog} \\
\hline
\end{tabular}
\end{table}
The nitrogen and seawater flow rates are determined by a normalized cooling command $u_{\mathrm{c}}$, where $0 \leq u_{\mathrm{c}} \leq 1$. After $u_{\mathrm{c}}$ is obtained, the equivalent number of active HX units, the rotational speed of each fan, and the seawater flow rate through each active heat-exchanger unit can be calculated.
The active fraction $f_{\mathrm{act}}$ is calculated as
\begin{equation}
f_{\mathrm{act}} = f_{\min} + \left( 1-f_{\min} \right) u_{\mathrm{c}}, \label{eq:active_cooling_fraction}
\end{equation}
where $f_{\min}$ is the minimum active fraction, kept nonzero so that a baseline level of cooling remains active at the minimum command. The active fraction increases continuously as $u_{\mathrm{c}}$ increases. The equivalent of active heat-exchanger units is calculated as 
\begin{equation} N_{\mathrm{HX,act}} = f_{\mathrm{act}}N_{\mathrm{HX}}, 
\label{eq:active_HX_units} 
\end{equation} 
where \(N_{\mathrm{HX}}\) is the total number of heat-exchanger units. Since \(f_{\mathrm{act}}\) is continuous, \(N_{\mathrm{HX,act}}\) represents an equivalent active HX count rather than an integer number of operating units in the subsea data center.

The rotational speed of each running fan is determined by 
\begin{equation} 
n_{\mathrm{fan}} = n_{\min} + u_{\mathrm{c}} \left( n_{\max}-n_{\min} \right), 
\label{eq:fan_rotational_speed} 
\end{equation} 
where \(n_{\min}\) and \(n_{\max}\) are the minimum and maximum fan rotational speeds, respectively. The volume flow rate of each single running fan is approximated using the fan affinity law~\cite{wu2014fan}, assuming that the volumetric flow rate scales linearly with fan rotational speed:
\begin{equation} 
\dot{V}_{\mathrm{fan}} = \dot{V}_{\mathrm{fan,rated}} \frac{n_{\mathrm{fan}}}{n_{\mathrm{rated}}}, \label{eq:fan_volume_flow} 
\end{equation} 
where $\dot{V}_{\mathrm{fan,rated}}$ and $n_{\mathrm{rated}}$ are the rated volume flow rate and rated rotational speed of the Delta FFB0412SHN fan~\cite{delta_ffb0412shn}, respectively. Here, $\dot{V}_{\mathrm{fan}}$ represents the volume flow rate of a single fan. The total circulating nitrogen volume flow rate is approximated by multiplying the single-fan volume flow rate by the equivalent number of active fans. Therefore,
\begin{equation}
\dot{m}_{\mathrm{N_2}} = f_{\mathrm{act}} N_{\mathrm{fan}} \rho_{\mathrm{N_2}} \dot{V}_{\mathrm{fan}}. 
\label{eq:N2_total_mass_flow1} 
\end{equation}
At the system level, the nitrogen flow is treated as the effective gas-side flow through the active HX path.


The seawater mass-flow rate through each active heat-exchanger unit is calculated as 
\begin{equation} 
\dot{m}_{\mathrm{sw,u}} = \dot{m}_{\mathrm{sw,u,min}} + u_{\mathrm{c}} \left( \dot{m}_{\mathrm{sw,u,max}} - \dot{m}_{\mathrm{sw,u,min}} \right), 
\label{eq:seawater_unit_mass_flow} 
\end{equation} where \(\dot{m}_{\mathrm{sw,u,min}}\) and \(\dot{m}_{\mathrm{sw,u,max}}\) 
are the minimum and maximum seawater mass-flow rates through one active heat-exchanger unit, respectively. The values listed in Table~\ref{tab:cooling_params} correspond to the lower and upper bounds of the seawater-side flow-rate range provided in the heat-exchanger effectiveness data used to determine $\varepsilon_{\mathrm{HX}}$~\cite{schmidt2005maintaining}. The total seawater mass-flow rate is then calculated as 
\begin{equation} 
\dot{m}_{\mathrm{sw}} = N_{\mathrm{HX,act}} \dot{m}_{\mathrm{sw,u}}. 
\label{eq:seawater_total_mass_flow} 
\end{equation} 
At \(u_{\mathrm{c}}=0\), the minimum fraction of the cooling equipment remains active, and each running unit operates at the minimum flow rate of the heat-exchanger effectiveness map. At \(u_{\mathrm{c}}=1\), all cooling units are active and operate at their maximum modeled flow rates. 

\subsection{Internal Nitrogen Circulation Fan Power} 
The internal circulation fans provide the nitrogen flow required for heat transfer between the IT equipment and the gas-to-seawater heat exchangers. The fan power is modeled as a function of the fan rotational speed. For each running fan, the power consumption is calculated using a polynomial approximation: 
\begin{equation} 
P_{\mathrm{fan,u}} = a_1 n_{\mathrm{fan}} + a_2 n_{\mathrm{fan}}^2 + a_3 n_{\mathrm{fan}}^3, 
\label{eq:fan_unit_power} 
\end{equation}
where \(P_{\mathrm{fan,u}}\) is the power consumption of one running fan, and \(n_{\mathrm{fan}}\) is the fan rotational speed in rpm obtained in~\eqref{eq:fan_rotational_speed}. The polynomial coefficients are adopted from the experimental fan-power model reported by Garraghan et al.~\cite{garraghan2016unified}, where Delta FFB0412SHN fans are tested to correlate fan rotational speed with electrical power consumption. The polynomial coefficients are set to $a_1 = 2.46\times 10^{-4}, \quad a_2 = -3.70\times 10^{-8}, \quad a_3 = 3.11\times 10^{-12}.$ Since only a fraction of the cooling stages may be active, the total fan power is calculated as 
\begin{equation} 
P_{\mathrm{fan}} = f_{\mathrm{act}}N_{\mathrm{fan}}P_{\mathrm{fan,u}}, \label{eq:fan_total_power} 
\end{equation} 
again \(f_{\mathrm{act}}\) is the active cooling fraction and \(N_{\mathrm{fan}}\) is the total number of circulation fans. 
\subsection{Seawater Pump Power}

The seawater pumps provide the external flow required by the gas-to-seawater heat exchangers. In this study, the detailed hydraulic pressure-loss model of the seawater loop is not explicitly resolved. Instead, a system-level variable-speed pump surrogate is used. Because the pump model only requires a reference flow rate and its corresponding reference power, the Calpeda BNMS 40/125A pump is selected as a representative pump. Other Calpeda pump models have also been used in literature \cite{oshurbekov2020energy,lawal2021energy}. The rated flow rate and power of the selected pump are adopted from the manufacturer’s catalog~\cite{calpeda_nms_catalog}.

The modeled cooling system contains \(N_{\mathrm{pump}}\) seawater pumps. Each
pump is assumed to supply one seawater branch. Similar to the HX, it also has an equivalent of active units computed as
\begin{equation}
N_{\mathrm{pump,act}}=f_{\mathrm{act}}N_{\mathrm{pump}}.
\end{equation}
To calculate the power consumption of a pump, the first step is to determine the seawater flow rate. Therefore, the seawater mass-flow rate handled by one running pump is calculated as
\begin{equation}
\dot{m}_{\mathrm{sw,pump}}
=\frac{N_{\mathrm{HX,act}}}{N_{\mathrm{pump,act}}}
\dot{m}_{\mathrm{sw},u}.
\label{eq:seawater_pump_mass_flow}
\end{equation}
Again \(\dot{m}_{\mathrm{sw},u}\) is the seawater mass-flow rate through one active heat-exchanger unit. To calculate the power consumption of the pump based on the centrifugal-pump affinity law~\cite{doe2007adjustable_speed_pumping}, we need to have the pump speed ratio $r_{\mathrm{pump}}$. After $\dot{m}_{\mathrm{sw,pump}}$ is obtained, it is converted to the volumetric flow $Q_{\mathrm{pump,u}}=3600 \dot{m}_{\mathrm{sw,pump}}/\rho_{\mathrm{sw}}$ and compared with reference flow $Q_{\mathrm{pump,ref}}$ of the Calpeda BNMS 40/125A pump. It gives the pump speed ratio $r_{\mathrm{pump}}=n_{\mathrm{pump}}/n_{\mathrm{pump,ref}}=Q_{\mathrm{pump,u}}/Q_{\mathrm{pump,ref}}$ using the centrifugal-pump affinity law $Q \propto n$. Again, using the centrifugal-pump affinity law $P \propto n^3 $~\cite{doe2007adjustable_speed_pumping}, the total seawater pump power is calculated as
\begin{equation}
P_{\mathrm{pump}}=f_{\mathrm{act}}N_{\mathrm{pump}}P_{\mathrm{pump,ref}}r_{\mathrm{pump}}^3.
\label{eq:pump_total_power}
\end{equation}
where $P_{\mathrm{pump,ref}}$ is the reference power of one pump when it runs at a reference flow rate. At $u_{\mathrm{c}}=1$, the modeled pump flow reaches the maximum heat-exchanger-side flow used in the simulation. At $u_{\mathrm{c}}=0$, only the minimum active cooling fraction remains in operation, and the seawater pump power decreases accordingly. Table~\ref{tab:cooling_params} lists the parameter values and their references used in this section.


\section{Energy Supply Modeling}
\label{sec-energy_modeling}
\begin{table}[t]
\centering
\caption{Energy-supply model parameters.}
\label{tab:energy_params}
\begin{tabular}{lll}
\hline
Parameter & Value & Source \\
\hline
\multicolumn{3}{l}{\textit{Wave power generation}}\\
$\eta_{\mathrm{wec}}$            & $0.40$                    & \cite{duah2025marine} \\
$g$                             & $9.8~\mathrm{m/s^2}$      & \cite{duah2025marine} \\
$L_{\mathrm{cap}}$              & $8.75~\mathrm{m}$         & \cite{dalton2010case} \\
$P_{\mathrm{wec,ref}}$          & $750~\mathrm{kW}$         & \cite{duah2025marine} \\
\hline
\multicolumn{3}{l}{\textit{Battery storage}}\\
$\eta_{\mathrm{ch}}$            & $0.9$                    & \cite{duah2025marine} \\
$\eta_{\mathrm{dis}}$           & $0.9$                    & \cite{duah2025marine} \\
$SOC_{\min}$                    & $0.10$                    & Simulation setting \\
$SOC_{\max}$                    & $1.00$                    & Simulation setting \\
$\Delta t$                      & $1~\mathrm{h}$            & Simulation setting \\
\hline
\end{tabular}
\end{table}
\subsection{Power Consumption Modeling}

The facility power demand is determined by the IT equipment power and cooling power. The IT power $P_{\mathrm{IT}}(t)$ is obtained from the workload model \eqref{eq:P_it} and is also used as the heat input to the thermal model. The heat-exchanger core itself is modeled as a passive thermal component. Therefore, it affects the thermal dynamics through $G_{\mathrm{HX}}$, but it does not introduce an additional electrical power term. The active cooling power consists of the internal nitrogen circulation fan power and the seawater pump power:
\begin{equation}
P_{\mathrm{cool}}(t)
=
P_{\mathrm{fan}}(t)
+
P_{\mathrm{pump}}(t).
\label{eq:P_cooling}
\end{equation}

The total data center load supplied by the wave-battery energy system is then calculated as
\begin{equation}
P_{\mathrm{load}}(t)
=
P_{\mathrm{IT}}(t)
+
P_{\mathrm{cool}}(t).
\label{eq:P_load}
\end{equation}
This load profile is used as the demand input for the renewable generation and battery storage model.

The corresponding IT energy consumption over the simulation horizon from $t_0$ to $t_f$ is calculated as
\begin{equation}
E_{\mathrm{IT}}
=
\int_{t_0}^{t_f}
P_{\mathrm{IT}}(t)\,dt.
\label{eq:E_IT}
\end{equation}
The active cooling energy consumption is calculated as
\begin{equation}
E_{\mathrm{cool}}
=
\int_{t_0}^{t_f}
P_{\mathrm{cool}}(t)\,dt.
\label{eq:E_cooling}
\end{equation}
The total load energy is therefore
\begin{equation}
E_{\mathrm{load}}
=
E_{\mathrm{IT}}
+
E_{\mathrm{cool}}.
\label{eq:E_load}
\end{equation}
Power usage effectiveness is used to evaluate the energy overhead of the active cooling system. In this study, it is calculated as $\mathrm{PUE}={E_{\mathrm{load}}}/{E_{\mathrm{IT}}}.$ This formulation directly measures the additional energy required by the active cooling system relative to the IT energy consumption.

\subsection{Wave Power Generation}

\begin{figure}[ptb]
    \centering
    \includegraphics[width=0.9\linewidth]{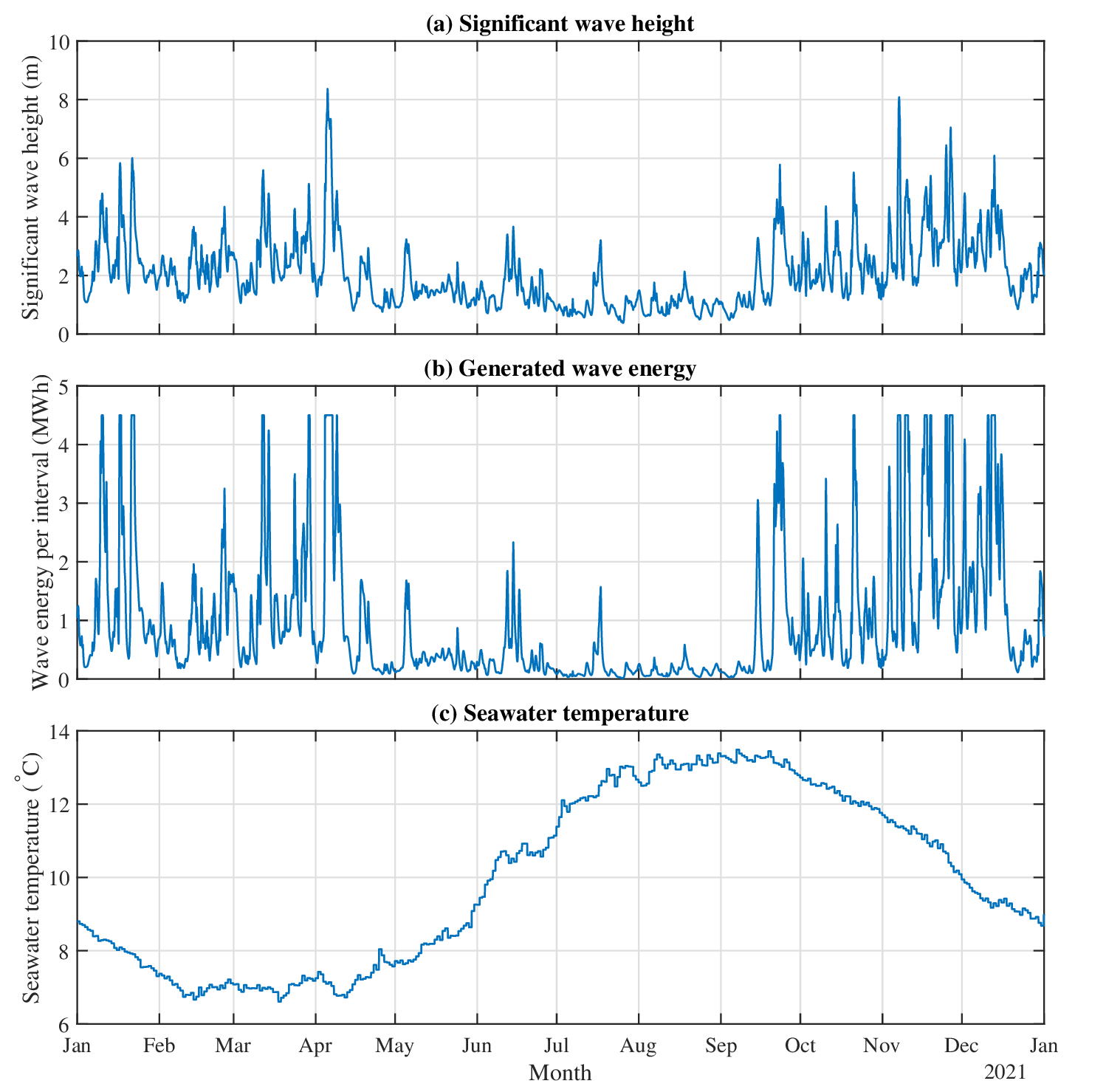}
    \caption{Full-year environmental and renewable-energy conditions.}
    \label{fig:wave_power_seatemp}
\end{figure}

Ocean wave energy is used as the renewable power source for the subsea data center. 
The electrical power generated by one wave energy converter is estimated from the deep-water wave power flux:
\begin{equation}
P_{\mathrm{wave,1,raw}}(t)=
\eta_{\mathrm{wec}}
\frac{\rho_{\mathrm{sw}}g^2}{32\pi}
H_{\mathrm{s}}^2(t)
T_{\mathrm{e}}(t)
L_{\mathrm{cap}},
\label{eq:single_wec_raw_power}
\end{equation}
where $H_{\mathrm{s}}(t)$ is the significant wave height, $T_{\mathrm{e}}(t)$ is the representative wave period, $\eta_{\mathrm{wec}}$ is the wave-to-electric conversion efficiency, $\rho_{\mathrm{sw}}$ is the seawater density, \(g\) is the gravitational acceleration, and $L_{\mathrm{cap}}$ is the effective capture width of one converter. This expression is based on the commonly used deep water wave energy flux formulation, in which the available wave power is proportional to the square of the significant wave height and linearly proportional to the wave period and capture width~\cite{duah2025marine,lenee2011characterizing,ozger2004stochastic}.


The electrical output of one wave energy converter is limited by its reference power:
\begin{equation}
P_{\mathrm{wave,1}}(t)=
\min
\left(
P_{\mathrm{wec,ref}},
P_{\mathrm{wave,1,raw}}(t)
\right),
\label{eq:single_wec_power_limit}
\end{equation}
where \(P_{\mathrm{wec,ref}}\) is the rated power of one converter. 
For an array with \(N_{\mathrm{wec}}\) identical converters, the total wave power is calculated as
\begin{equation}
P_{\mathrm{wave}}(t)
=
N_{\mathrm{wec}}P_{\mathrm{wave,1}}(t). 
\label{eq:total_wave_power}
\end{equation}
The resulting \(P_{\mathrm{wave}}(t)\) is then used together with the data center load profile in the battery storage model. 

In this study, the wave resource, i.e., the significant wave height $H_{\mathrm{s}}(t)$ and the mean wave period $T_{\mathrm{e}}(t)$, is obtained from ERA5 reanalysis data~\cite{hersbach2020era5}. 
Figure~\ref{fig:wave_power_seatemp} shows the annual variations in significant wave height, generated wave energy, and seawater temperature over one year. For this particular dataset, wave activity and energy generation are generally higher in winter and lower in summer, while seawater temperature follows a clear seasonal trend.

\subsection{Battery Storage System}

A battery energy storage system is used to buffer the mismatch between wave power generation and data center load. At each time step, the net power balance is calculated as
\begin{equation}
P_{\mathrm{net}}(k)=P_{\mathrm{wave}}(k)-P_{\mathrm{load}}(k),
\label{eq:net_power}
\end{equation}
where \(P_{\mathrm{wave}}(k)\) is the total wave power generation and \(P_{\mathrm{load}}(k)\) is the data center load including IT power and active cooling power. 

When \(P_{\mathrm{net}}(k)\geq 0\), the wave generation is larger than the data center load. The surplus power is used to charge the battery. Based on \cite{lei2020dynamic}, the battery SOC during charging is updated as
\begin{equation}
SOC(k+1) = SOC(k)+
\frac{P_{\mathrm{ch}}(k)\Delta t \eta_{\mathrm{ch}}}{E_{\mathrm{bat}}}, 
\label{eq:SOC_charge}
\end{equation}
where $E_{\mathrm{bat}}$ is the battery storage capacity. The charging power is subject to the battery power limit and the remaining available energy capacity:
\begin{equation} 
P_{\mathrm{ch}}(k) = \min \left[ \begin{aligned} & P_{\mathrm{net}}(k), \\ & P_{\mathrm{bat}}, \\ & \frac{ \left(SOC_{\max}-SOC(k)\right)E_{\mathrm{bat}} } {\Delta t \eta_{\mathrm{ch}}} 
\end{aligned} \right], 
\label{eq:battery_charge_power} 
\end{equation}
where \(P_{\mathrm{bat}}\) is the maximum battery charging and discharging power, \(SOC_{\max}\) is the maximum state of charge, and \(\eta_{\mathrm{ch}}\) is the charging efficiency. The third term represents the maximum allowable charging power determined by the remaining battery capacity, ensuring that the SOC does not exceed \(SOC_{\max}\) over the time step \(\Delta t\) while accounting for the charging efficiency. 


When \(P_{\mathrm{net}}(k)<0\), the wave generation is lower than the data center load. 
The battery discharges to compensate for this deficit. Similar to the battery charging~\cite{lei2020dynamic}, the battery SOC during discharging is updated as
\begin{equation}
SOC(k+1)=SOC(k)-\frac{P_{\mathrm{dis}}(k)\Delta t}
{\eta_{\mathrm{dis}}E_{\mathrm{bat}}}.
\label{eq:SOC_discharge}
\end{equation}
The actual discharge power is limited by the load deficit, the battery power capacity, and the available stored energy above the minimum SOC:
\begin{equation}
P_{\mathrm{dis}}(k)=\min\left[
\begin{aligned}
& -P_{\mathrm{net}}(k), \\
& P_{\mathrm{bat}}, \\
& \frac{
\left(SOC(k)-SOC_{\min}\right)
E_{\mathrm{bat}}\eta_{\mathrm{dis}}}{\Delta t}
\end{aligned}
\right],
\label{eq:battery_discharge_power}
\end{equation}
The third term represents the maximum allowable discharging power determined by the available battery energy above $SOC_{\min}$, ensuring that the SOC does not fall below its minimum limit over the time step $\Delta t$ while accounting for the discharging efficiency. 

Table~\ref{tab:energy_params} lists the parameter values and their references used in this section.

\section{NMPC-Based Cooling and Workload Control}\label{sec-optimization}

In a wave-powered subsea data center, the available renewable power is time-varying, while the flexible workload can be delayed~\cite{han2023two}. Therefore, executing flexible jobs only according to their arrival times or regulating cooling only according to the current temperature may lead to excessive cooling energy consumption or delayed flexible-workload execution. To address these issues, an NMPC formulation is used to coordinate flexible workload execution and active cooling operation. The controller predicts the future wave generation, interactive workload, flexible-workload queue, seawater temperature, battery SOC, and thermal states over a finite prediction horizon. At each time step, the controller optimizes the flexible workload power budget and the normalized cooling command.

At time step $k$, the predicted IT power at prediction step $j$, based on \eqref{eq:P_it}, is calculated as
\begin{equation}
P_{\mathrm{IT},j}=P_{\mathrm{base}}+P_{\mathrm{interactive},j}+P_{\mathrm{flex},j},
\end{equation} 
where the base IT power and the interactive workload power \(P_{\mathrm{interactive},j}\) are treated as known mandatory workload components. The controller then determines the total power budget $P_{\mathrm{flex},j}$ allocated to flexible workloads and the cooling command $u_{\mathrm{c},j}$ over the prediction horizon. When the predicted wave power and battery SOC are sufficient, the controller can increase $P_{\mathrm{flex},j}$ to reduce the waiting queue. When the available power is low, the battery SOC is close to its reserve level, or the IT temperature is high, the controller reduces $P_{\mathrm{flex},j}$, delays part of the flexible workload and only executes committed flexible jobs. At the same time, $u_{\mathrm{c},j}$ is adjusted to keep the IT temperature below its upper limit while avoiding unnecessary fan and pump power. Only the first cooling command $u_{\mathrm{c},0}$ is directly applied to the cooling system. The first optimized flexible-workload power budget $P_{\mathrm{flex},0}$ is provided to the strict first-come-first-served (FCFS) job-level scheduler, which determines the jobs admitted during the current sampling interval. Because the workload consists of discrete jobs, the actual executed flexible-workload power may be lower than $P_{\mathrm{flex},0}$. The optimization is repeated at the next sampling time using the updated system states, committed-job power and flexible-workload queue.

The prediction horizon is denoted by $N_p$.
At each sampling time $k$, the decision variables are the flexible-workload power sequence
\begin{equation}
\mathbf{P}_{\mathrm{flex},k}
=
\left\{
P_{\mathrm{flex},0},
P_{\mathrm{flex},1},
\dots,
P_{\mathrm{flex},N_p-1}
\right\},
\label{eq:flex_power_sequence}
\end{equation}
and the cooling-command sequence
\begin{equation}
\mathbf{U}_{\mathrm{c},k}
=
\left\{
u_{\mathrm{c},0},
u_{\mathrm{c},1},
\dots,
u_{\mathrm{c},N_p-1}
\right\},
\label{eq:cooling_command_sequence}
\end{equation}
where $j=0,\ldots,N_p-1$ denotes the local prediction-step index within the NMPC horizon.
The normalized cooling command satisfies $u_{\mathrm{c},j}\in[0,1]$, which determines the active cooling fraction, fan speed, seawater flow rate, fan power, and pump power according to the cooling model defined previously in Section~\ref{secsub-cooling_control}.

To prevent flexible jobs from accumulating in the waiting queue, the remaining flexible workload is represented in energy form and included as an additional state in the NMPC model. Its evolution is given by
\begin{equation}
Q_{\mathrm{flex},j+1}=Q_{\mathrm{flex},j}+E_{\mathrm{arr},j}-
\left(P_{\mathrm{flex},j}-P_{\mathrm{comm},j}\right)\Delta t,
\label{eq:flex_queue_update}
\end{equation}
where $Q_{\mathrm{flex},j}$ is the total remaining energy of the flexible jobs waiting in the queue at prediction step $j$. The term $E_{\mathrm{arr},j}$ denotes the energy of the flexible jobs arriving during that prediction interval and is obtained directly from the workload data described in Section~\ref{secsub-IT_workload}. The term $P_{\mathrm{flex},j}-P_{\mathrm{comm},j}$ represents the planned power available for admitting additional waiting flexible jobs. Multiplying this value by $\Delta t$ gives the energy that can be removed during the prediction interval. Because actual job admission is performed by the strict FCFS scheduler, the actual reduction in queue energy may be smaller than this predicted value.


The predicted state vector is written as
\begin{equation}
\mathbf{x}_j
=
\begin{bmatrix}
T_{\mathrm{IT},j} &
T_{\mathrm{N_2},j} &
T_{\mathrm{h},j} &
SOC_j &
Q_{\mathrm{flex},j}
\end{bmatrix}^{\top},
\label{eq:nmpc_state_vector}
\end{equation}

At each sampling step $k$, the controller solves the following optimal control problem:
\begin{subequations}
\label{eq:nmpc_formulation}
\begin{align}
\min_{\mathbf{P}_{\mathrm{flex},k},\,\mathbf{U}_{\mathrm{c},k}}
\quad
&
J=
\sum_{j=0}^{N_p-1}
\ell_j
+
V_f
\left(
\mathbf{x}_{N_p}
\right)\\
\text{s.t.}\quad&
\mathbf{x}_{j+1}=f_{\mathrm{pred}}
\left(
\mathbf{x}_j,
P_{\mathrm{flex},j},
u_{\mathrm{c},j}
\right),\\&
P_{\mathrm{comm},j}\leq
P_{\mathrm{flex},j}\leq P_{\mathrm{flex,max}},
\quad
0\leq u_{\mathrm{c},j}\leq1,\\&
\hspace{20mm}j=0,\ldots,N_p-1,\\&
\mathbf{x}_0=\mathbf{x}(k),\\&
SOC_{\min}\leq SOC_j\leq SOC_{\max},\\&
T_{\mathrm{IT},j}\leq T_{\mathrm{IT,max}}.
\end{align}
\end{subequations}
Here, \(k\) denotes the current sampling step, whereas \(j\) denotes the local prediction-step index within the NMPC horizon. At each sampling step, the NMPC determines the flexible-workload power sequence $\mathbf{P}_{\mathrm{flex},k}$ and the cooling-command sequence $\mathbf{U}_{\mathrm{c},k}$ over the prediction horizon. The objective function consists of the stage cost $\ell_j$ (defined in \eqref{eq:nmpc_stage_cost} below), accumulated over the prediction horizon, and the terminal cost $V_f(\mathbf{x}_{N_p})$ (defined in \eqref{eq:nmpc_terminal_cost} below). The prediction model $f_{\mathrm{pred}}(\cdot)$ combines the discretized three-node thermal balances in \eqref{eq:IT_balance}-\eqref{eq:Q_h_sw}, the battery charging, discharging, and SOC-update relations in \eqref{eq:net_power}-\eqref{eq:battery_discharge_power}, and the flexible-workload queue update in \eqref{eq:flex_queue_update}. The constraints ensure that the flexible-workload power is no lower than the power required by previously committed jobs, limit the flexible-workload power and cooling command to their allowable ranges, maintain the battery SOC within its operating limits, and keep the IT temperature below its maximum allowable value. When the measured SOC is at or below the new-flexible-job admission threshold, the first-step flexible-workload power is restricted to the power required by previously committed jobs:
\begin{equation}
P_{\mathrm{flex},0}=P_{\mathrm{comm},0},
\qquad
\text{if } SOC(k)\leq SOC_{\mathrm{stop}}.
\label{eq:new_flex_soc_stop}
\end{equation}
Therefore, no new flexible job is admitted, while previously committed jobs continue running.

The stage cost is defined as
\begin{equation}
\begin{aligned}
\ell_j={}&w_{\mathrm{q}}Q_{\mathrm{flex},j+1}^{2}+w_{\mathrm{cool}}
u_{\mathrm{c},j}^{2}\\&+
w_{\mathrm{SOC}}
\left[\max\left(0,SOC_{\mathrm{target}}-SOC_{j+1}\right)\right]^2
\\&+
w_{\mathrm{T}}\left[\max\left(0,
T_{\mathrm{IT},j+1}-T_{\mathrm{guard}}\right)\right]^2
\\&+w_{\Delta u}\left(\Delta u_{\mathrm{c},j}\right)^2+w_{\Delta P}
\left(\Delta P_{\mathrm{flex},j}\right)^2 .
\end{aligned}
\label{eq:nmpc_stage_cost}
\end{equation}
The queue term penalizes the remaining flexible-workload backlog. The second term discourages high cooling commands. The SOC penalty is activated when the predicted SOC falls below $SOC_{\mathrm{target}}$. The temperature term penalizes operation within the guard region near the IT-temperature limit. The final two terms penalize variations in the predicted cooling command and flexible-workload power, respectively. 

The temperature guard threshold is defined as
\begin{equation}
T_{\mathrm{guard}}=T_{\mathrm{IT,max}}-\Delta T_{\mathrm{guard}},
\label{eq:temperature_guard}
\end{equation}
where \(\Delta T_{\mathrm{guard}}\) is the prescribed temperature guard margin.

The terminal cost is defined as
\begin{equation}
\begin{aligned}
V_f\left(\mathbf{x}_{N_p}\right)={}&
w_{\mathrm{SOC},f}\left[\max\left(0,SOC_{\mathrm{target}}-SOC_{N_p}\right)
\right]^2\\&+w_{\mathrm{q},f}Q_{\mathrm{flex},N_p}^{2},
\end{aligned}
\label{eq:nmpc_terminal_cost}
\end{equation}
where $w_{\mathrm{SOC},f}$ and $w_{\mathrm{q},f}$ are the terminal SOC and terminal queue weights, respectively. The terminal cost encourages the controller to preserve the battery reserve and reduce the remaining flexible-workload backlog at the end of the prediction horizon.

The optimized flexible-workload power is used as an admission budget for queued flexible jobs. The actual flexible-job execution follows a strict FCFS rule, so the controller does not select individual flexible jobs out of order. Only the first-step optimized power budget is used for job admission because the NMPC problem is solved again at the next sampling time. Once admitted, a flexible job remains committed until completed. A new job is admitted only if its current-step power satisfies the available flexible-workload power budget and sufficient battery energy remains available above the flexible-job reserve threshold $SOC_{\mathrm{flex}}$ for the remaining execution of all committed jobs. If the oldest waiting job cannot satisfy these conditions, the admission process stops and no subsequent job is allowed to bypass it. Unlike $SOC_{\mathrm{flex}}$, which preserves sufficient battery energy for future flexible-job execution, $SOC_{\mathrm{stop}}$ acts as a hard admission threshold: when the measured SOC is at or below $SOC_{\mathrm{stop}}$, no new flexible job is admitted, while previously committed jobs continue running.

\section{Results and Discussion}\label{sec-results}
Several parametric studies are conducted to evaluate the proposed control framework under different workload, environmental, thermal, and system-design conditions. For all cases, the same three-node thermal model in \eqref{eq:IT_balance}--\eqref{eq:hull_balance}, wave-battery energy model in \eqref{eq:single_wec_raw_power}--\eqref{eq:total_wave_power}, strict FCFS workload-admission rule, and NMPC formulation in \eqref{eq:nmpc_formulation}--\eqref{eq:nmpc_terminal_cost} are used. 

All closed-loop simulations are implemented in MATLAB using a control sampling interval of $\Delta t=1~\mathrm{h}$. At each sampling time step, one NMPC problem is solved using the SQP algorithm implemented in fmincon. The prediction horizon is set to $N_{p}=8$, corresponding to an $8$-h prediction window. Lastly, $D_{\max}=48~\mathrm{h}$ is adopted as the baseline deadline for the following studies. 

Each workload case is simulated for nine days. Flexible jobs arriving during the first seven days are included in the job-level QoS metrics, whereas the final two days allow jobs already present in the system to continue affecting the physical workload, battery, thermal states, and FCFS queue. The physical minimum SOC $SOC_{\min}$, flexible-job reserve threshold $SOC_{\mathrm{flex}}$, new-job admission threshold $SOC_{\mathrm{stop}}$, and operating SOC target $SOC_{\mathrm{target}}$ are set to 0.10, 0.20, 0.40, and 0.55, respectively. Only the investigated parameter or input dataset is changed in each study. The weighting coefficients used in the NMPC objective function are summarized in Table~\ref{tab:nmpc_weights}.
\begin{table}[t]
\centering
\caption{NMPC weighting coefficients.}
\label{tab:nmpc_weights}
\small
\setlength{\tabcolsep}{3.5pt}
\begin{tabular}{l c c}
\hline
Cost term & Weight & Value \\
\hline
Flexible-workload backlog & $w_{\mathrm{q}}$ & $3.0\times10^{-17}$ \\
Cooling command & $w_{\mathrm{cool}}$ & $0.30$ \\
Low-SOC penalty & $w_{\mathrm{SOC}}$ & $1000$ \\
Temperature-guard penalty & $w_{\mathrm{T}}$ & $20$ \\
Cooling-command variation & $w_{\Delta u}$ & $0.10$ \\
Flexible-power variation & $w_{\Delta P}$ & $5.0\times10^{-10}$ \\
Terminal SOC penalty & $w_{\mathrm{SOC},f}$ & $2500$ \\
Terminal backlog penalty & $w_{\mathrm{q},f}$ & $6.0\times10^{-17}$ \\
\hline
\end{tabular}
\end{table}

\subsection{Impacts of Temperature Limit $T_{\mathrm{IT,max}}$}

\begin{figure}[ptb]
    \centering
    \includegraphics[width=0.9\linewidth]{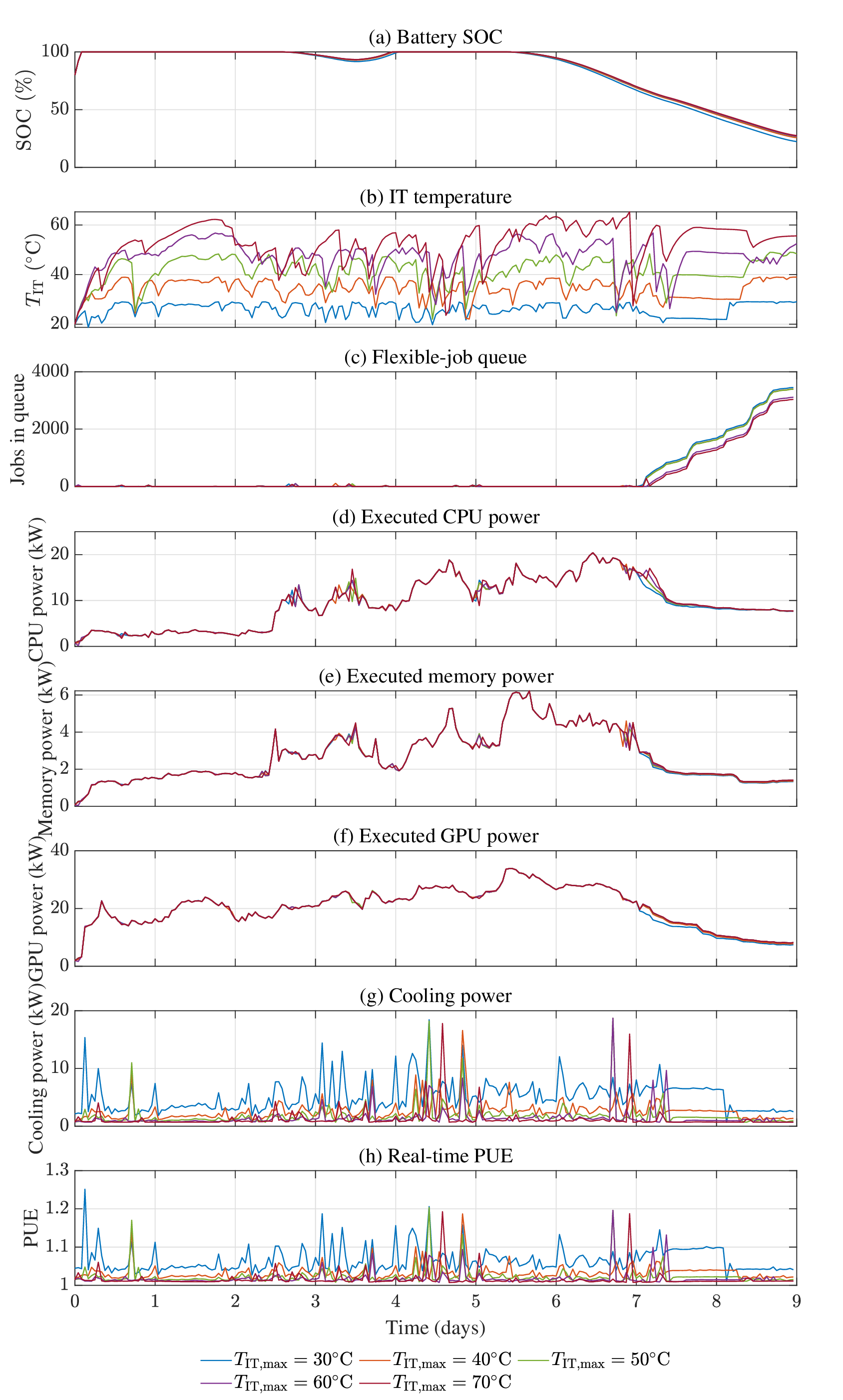}
    \caption{Closed-loop trajectories under different IT upper temperature limits.}
    \label{fig:temperature_sensitivity_eight_panel}
\end{figure}
\begin{table}[ptb]
\centering
\caption{Flexible-job QoS under different IT upper-temperature limits.}
\label{tab:temperature_sensitivity_qos}
\resizebox{\columnwidth}{!}{
\begin{tabular}{c c c c c c}
\hline
\begin{tabular}[c]{@{}c@{}}
$T_{\mathrm{IT,max}}$\\
($^{\circ}$C)
\end{tabular}&\begin{tabular}[c]{@{}c@{}}Delayed\\jobs(\%)
\end{tabular}&\begin{tabular}[c]{@{}c@{}}Mean\\
delay (h)\end{tabular}&\begin{tabular}[c]{@{}c@{}}P90\\
delay (h)\end{tabular}&\begin{tabular}[c]{@{}c@{}}Maximum\\
delay (h)\end{tabular}&\begin{tabular}[c]{@{}c@{}}Missed\\
jobs\end{tabular}\\
\hline
30 & 3.097 & 1.063 & 1 & 2 & 0 \\
40 & 3.702 & 1.015 & 1 & 2 & 0 \\
50 & 3.014 & 1.000 & 1 & 1 & 0 \\
60 & 3.091 & 1.000 & 1 & 1 & 0 \\
70 & 3.257 & 1.028 & 1 & 2 & 0 \\
\hline
\end{tabular}
}
\end{table}
The first study evaluates the effect of the maximum allowable IT temperature $T_{\mathrm{IT,max}}$. The temperature limit is varied over $T_{\mathrm{IT,max}}\in\{30,40,50,60,70\}^{\circ}\mathrm{C}$. This study examines the tradeoff among thermal constraint tightness, cooling demand, IT temperature, battery operation, PUE, and flexible-workload scheduling. Unless otherwise stated, the baseline configuration uses the workload window beginning on \textit{Jul.~17}, $N_{\mathrm{wec}}=6$, $E_{\mathrm{bat}}=6~\mathrm{MWh}$, $P_{\mathrm{bat}}=0.8~\mathrm{MW}$, $P_{\mathrm{base}}=45~\mathrm{kW}$, $T_{\mathrm{IT,max}}=40^{\circ}\mathrm{C}$, and $D_{\max}=48~\mathrm{h}$. 

Figure~\ref{fig:temperature_sensitivity_eight_panel} shows the corresponding closed-loop trajectories. As the allowable temperature increases, the IT equipment generally operates at a higher temperature, while the cooling power and real-time PUE are reduced. The battery SOC trajectories remain similar during most of the simulation, although the cases with higher temperature limits retain slightly more energy near the end. The executed CPU, memory, and GPU power trajectories are also similar across the evaluated cases. More noticeable differences appear near the end of the simulation, when the flexible-job queues begin to increase.

The job-level quality of service (QoS) results are summarized in Table~\ref{tab:temperature_sensitivity_qos}. A delayed job is a flexible job that is not started immediately upon arrival. The delayed-job percentage includes both delayed jobs that eventually start and jobs that remain unstarted at the end of the simulation. The mean, P90, and maximum delays are calculated only for delayed jobs that eventually start, excluding jobs that remain unstarted at the end of the simulation. A missed job is one whose waiting time exceeds the prescribed deadline. The delayed-job percentage varies slightly among the temperature-limit cases without showing a consistent trend. The mean and P90 delays remain similar, while the maximum delay is limited to a small number of hours. No missed jobs are observed in any evaluated case.

Overall, increasing the IT temperature limit reduces cooling demand and slightly improves the battery SOC, but it does not consistently improve flexible-job QoS. The $40^{\circ}\mathrm{C}$ case is therefore retained as the baseline for the subsequent parametric studies because it provides a moderate temperature limit while maintaining acceptable thermal and workload-scheduling performance.



\subsection{Impacts of Week of Year}
\begin{figure}[ptb]
    \centering
    \includegraphics[width=0.9\linewidth]{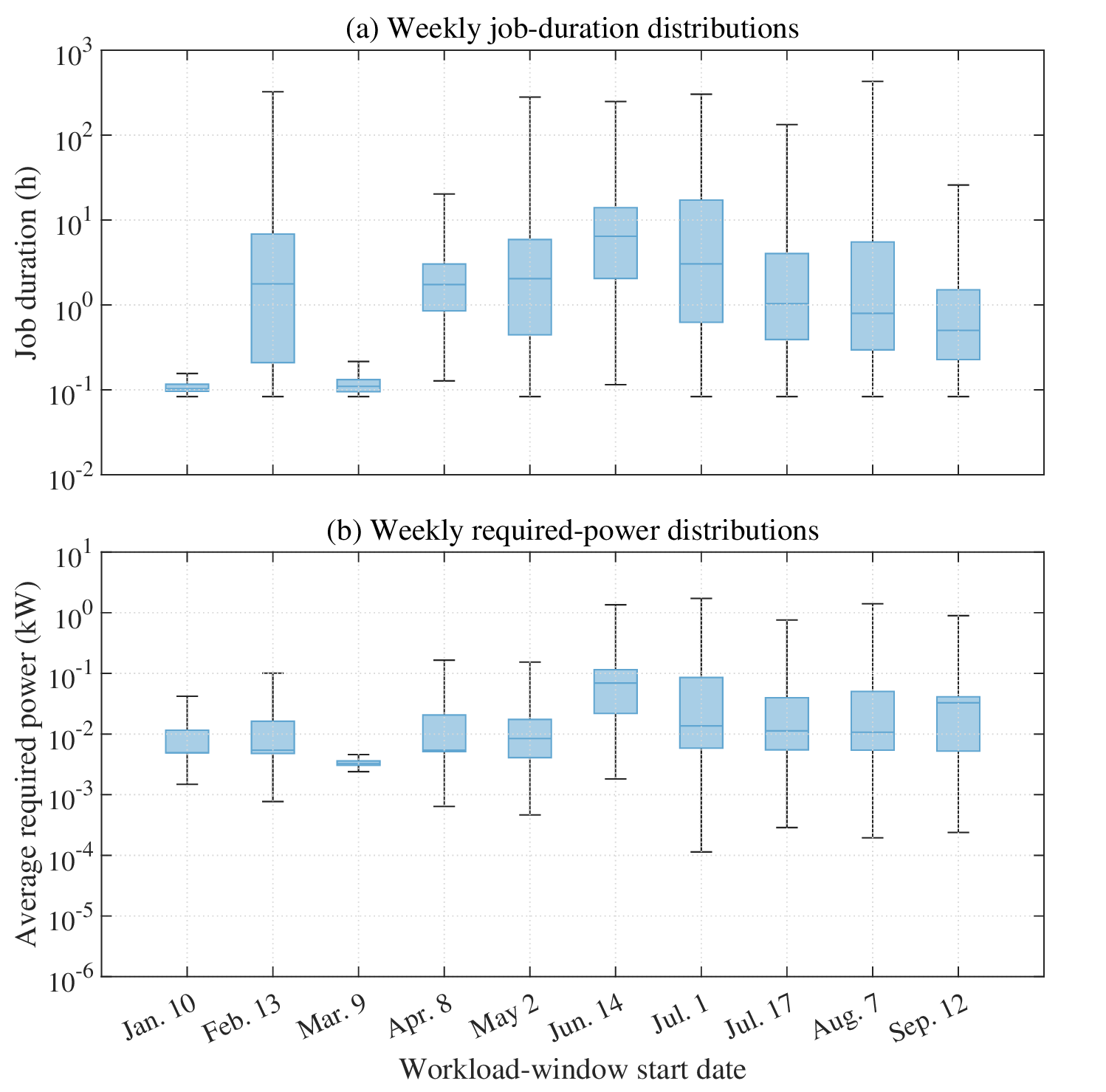}
    \caption{Weekly distributions of flexible-job characteristics.}
    \label{fig:weekly_job_distribution}
\end{figure}

\begin{figure}[ptb]
    \centering
    \includegraphics[width=0.9\linewidth]{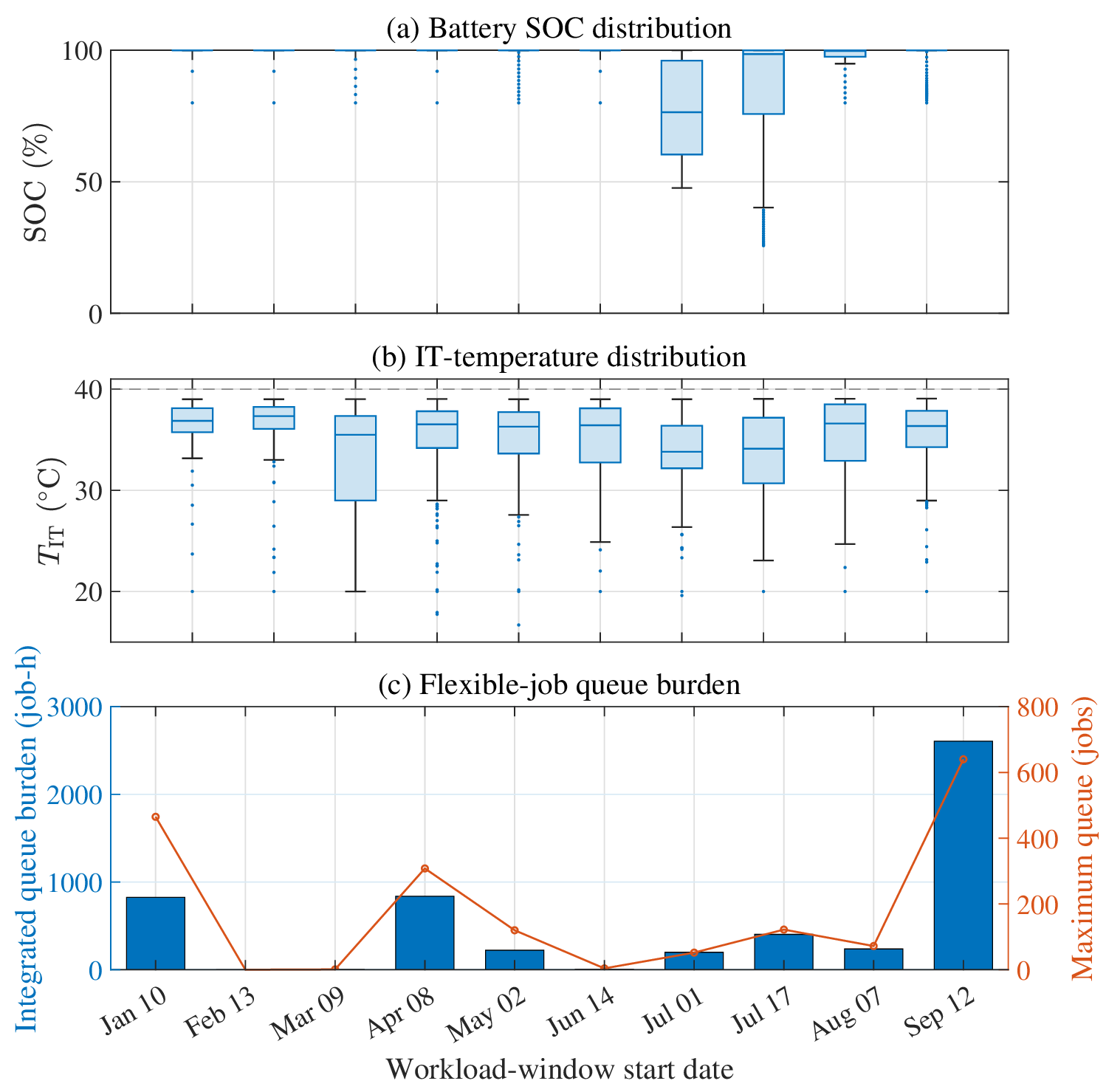}
    \caption{Distributions of closed-loop operating variables under different weekly workload and environmental conditions.} 
    \label{fig:week_sensitivity_boxplot_nofilter}
\end{figure}
The second study evaluates the closed-loop performance under different weekly workload and environmental conditions. Ten workload windows distributed across the available dataset are selected to capture variations in flexible-job arrivals, wave power availability, and seawater temperature. The comparison is used to evaluate the effects of weekly operating conditions on battery operation, cooling demand, queue accumulation, and job-level QoS. In this study, the IT upper-temperature limit is fixed at $40^{\circ}\mathrm{C}$, and the remaining system and controller settings are unchanged.

Figure~\ref{fig:weekly_job_distribution} shows substantial week-to-week variation in the flexible-job characteristics. The \textit{Jan.~10} and \textit{Mar.~9} windows are dominated by relatively short and low-power flexible jobs, whereas the \textit{Feb.~13} and \textit{Jul.~1} windows have broader job duration and required power distributions. These differences indicate that the evaluated weekly cases impose different scheduling and energy management demands on the system.

The simulation results are shown in Fig.~\ref{fig:week_sensitivity_boxplot_nofilter}. The integrated queue burden is measured in job-hours, where one job-hour represents one flexible job remaining in the queue for one hour. Therefore, this metric accounts for both the number of waiting jobs and the duration of queue accumulation. The battery SOC remains close to its upper range for most weeks, while wider SOC distributions and larger reductions are observed for \textit{Jul.~1} and \textit{Jul.~17}. The IT temperature remains below the $40^{\circ}\mathrm{C}$ limit in all cases, although its distribution varies across the workload weeks. \textit{Sep.~12} exhibits the largest integrated queue burden and maximum queue size, indicating the most severe and heavy queue accumulation. \textit{Jan.~10} and \textit{Apr.~8} also show noticeable queue accumulation, whereas the remaining weekly cases have relatively limited queue burdens. 

The corresponding QoS results are summarized in Table~\ref{tab:week_sensitivity_all_jobs}, showing clear differences in flexible-job delay performance across the weekly cases. Although \textit{Sep.~12} exhibits the largest integrated queue burden and maximum queue size, its delayed-job percentage is not the highest among the evaluated weeks, indicating that more severe queue accumulation does not necessarily correspond to a larger fraction of delayed jobs. Overall, the results demonstrate that variations in workload characteristics and environmental conditions affect battery utilization, queue accumulation, and flexible-job service performance.


\begin{table}[ptb]
\centering
\caption{Flexible-job QoS across different weeks.}
\label{tab:week_sensitivity_all_jobs}
\resizebox{\columnwidth}{!}{
\begin{tabular}{c c c c c c}
\hline
\begin{tabular}[c]{@{}c@{}}Week\\ start\end{tabular} &
\begin{tabular}[c]{@{}c@{}}Delayed\\ jobs (\%)\end{tabular} &
\begin{tabular}[c]{@{}c@{}}Mean\\ delay (h)\end{tabular} &
\begin{tabular}[c]{@{}c@{}}P90\\ delay (h)\end{tabular} &
\begin{tabular}[c]{@{}c@{}}Maximum\\ delay (h)\end{tabular} &
\begin{tabular}[c]{@{}c@{}}Missed\\ jobs\end{tabular} \\
\hline
\textit{Jan. 10} & 7.93 & 1.77 & 2.00 & 3 & 0 \\
\textit{Feb. 13} & 0.00 & 0.00 & 0.00 & 0 & 0 \\
\textit{Mar. 9} & 0.05 & 1.00 & 1.00 & 1 & 0 \\
\textit{Apr. 8}  & 8.28 & 1.29 & 2.00 & 2 & 0 \\
\textit{May 2 }  & 4.99 & 1.00 & 1.00 & 1 & 0 \\
\textit{Jun. 14} & 0.17 & 1.00 & 1.00 & 1 & 0 \\
\textit{Jul. 1}  & 3.37 & 1.29 & 2.00 & 2 & 0 \\
\textit{Jul. 17} & 2.55 & 1.10 & 1.00 & 2 & 0 \\
\textit{Aug. 7}  & 2.10 & 1.59 & 3.00 & 6 & 0 \\
\textit{Sep. 12} & 4.79 & 2.05 & 5.00 & 7 & 0 \\
\hline
\end{tabular}
}
\end{table}



\begin{figure}[ptb]
    \centering
    \includegraphics[width=0.9\linewidth]{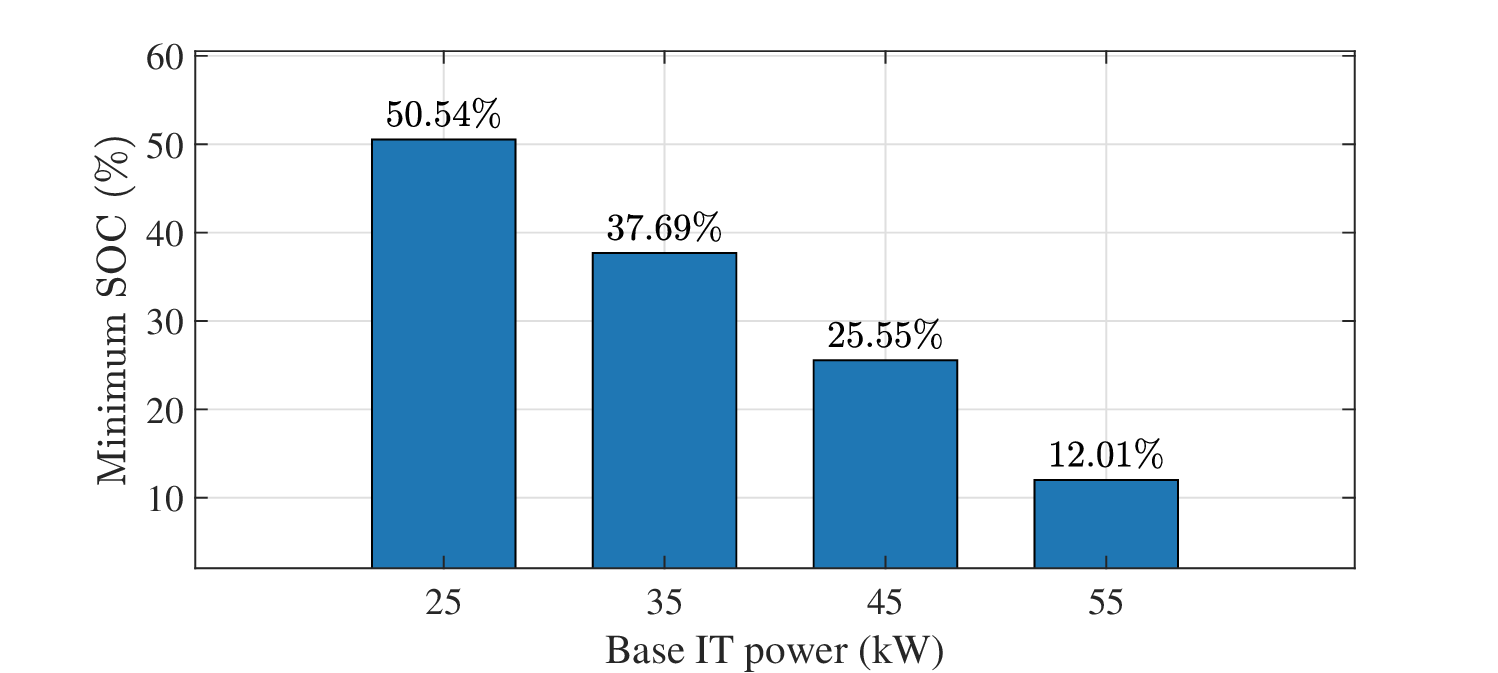}
    \caption{Minimum battery SOC under different base IT power.}
    \label{fig:base_power_sensitivity_minimum_SOC}
\end{figure}

\begin{figure}[ptb]
    \centering
    \includegraphics[width=0.9\linewidth]{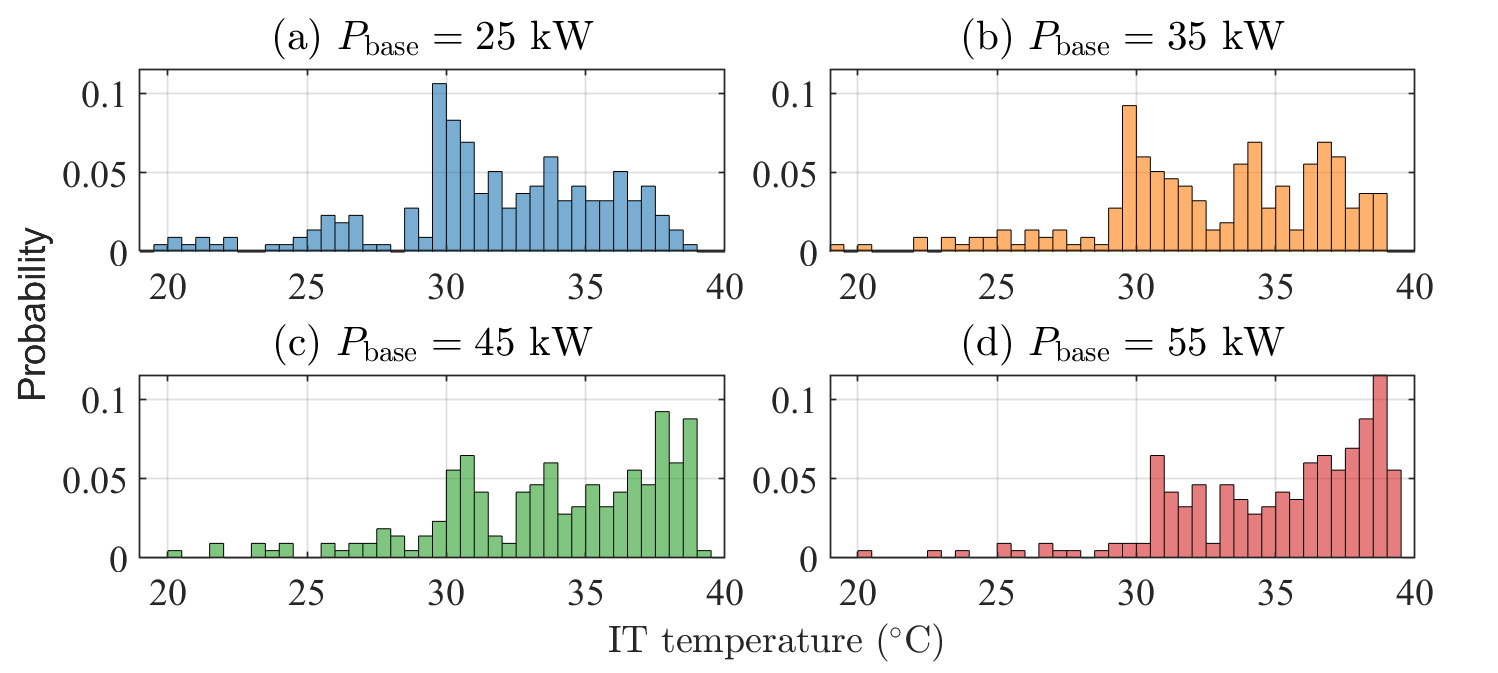}
    \caption{IT-Temperature distributions under different base IT power.}
    \label{fig:base_power_sensitivity_temp}
\end{figure}

\begin{figure}[ptb]
    \centering
    \includegraphics[width=0.9\linewidth]{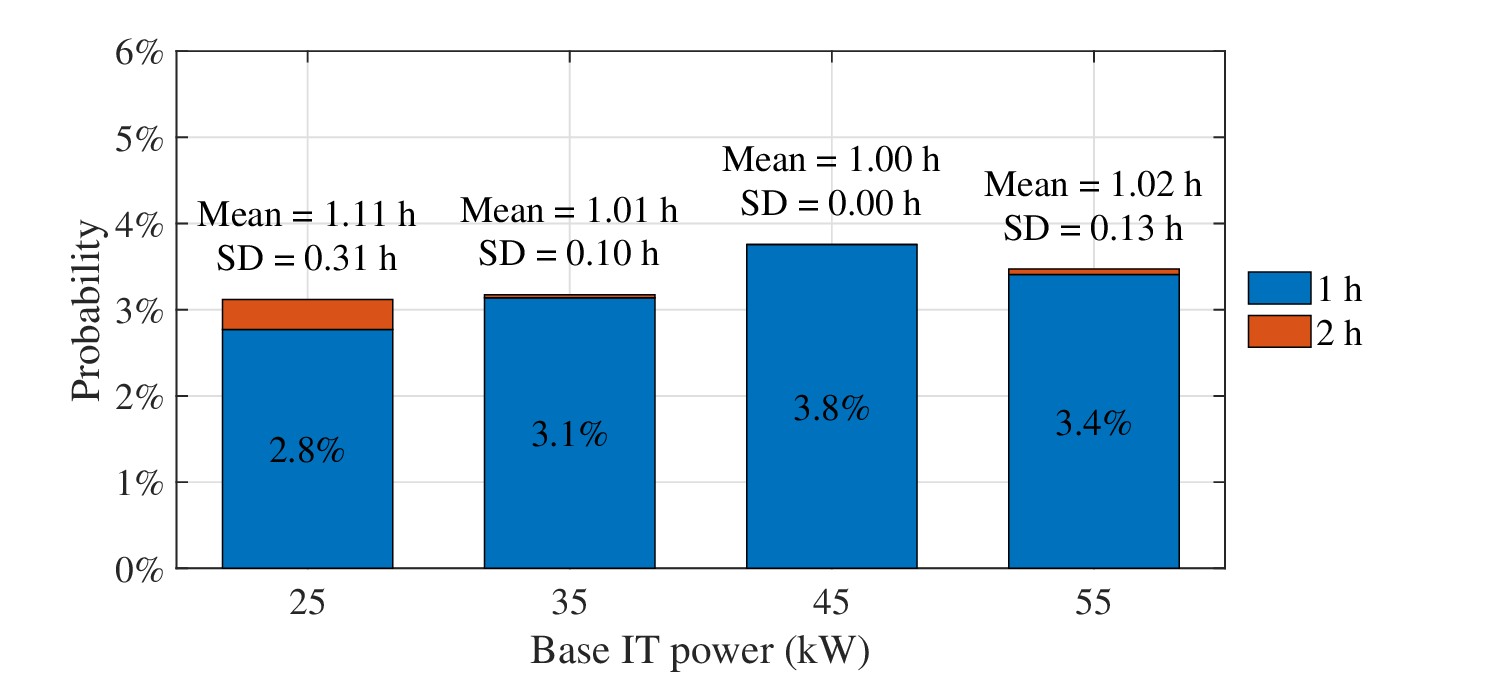}
    \caption{Flexible-Job delay distributions under different base IT power.}
    \label{fig:base_power_sensitivity_job}
\end{figure}


\subsection{Impacts of Base-IT-Power}
The fourth study evaluates the influence of the continuously operating base IT power. With the IT upper-temperature limit fixed at $40^{\circ}\mathrm{C}$, the base power is varied over $P_{\mathrm{base}}\in\{25,35,45,55\}~\mathrm{kW}$. The workload, environmental conditions, battery configuration, and controller settings remain unchanged, and the week beginning on July 17 is evaluated in all cases.


As shown in Figs.~\ref{fig:base_power_sensitivity_minimum_SOC}--\ref{fig:base_power_sensitivity_job}, increasing the base IT power consistently reduces the minimum battery SOC and shifts the IT temperature distribution toward higher temperatures. However, the IT temperature remains below its upper limit in all evaluated cases. The corresponding QoS results in Table~\ref{tab:base_power_sensitivity_qos} show only minor variations across the evaluated cases, with $3.1\%$--$3.8\%$ of the started flexible jobs experiencing delays and no flexible job missing its deadline. All delayed jobs wait for either one or two hours, with mean delays of $1.00$--$1.11~\mathrm{h}$ and standard deviations no greater than $0.31~\mathrm{h}$. No clear monotonic trend in QoS is observed as the base IT power increases. Table~\ref{tab:base_power_sensitivity_pue} further shows that the PUE remains nearly unchanged across the evaluated base-power settings.

Overall, increasing the base IT power mainly reduces the available battery-energy margin and raises the IT operating temperature, while producing only minor changes in flexible-job QoS and PUE. The $45~\mathrm{kW}$ case is retained as the nominal setting for the other studies.

\begin{table}[ptb]
\centering
\caption{Flexible-job QoS under different base IT powers.}
\label{tab:base_power_sensitivity_qos}
\small
\begin{tabular}{c c c c c c}
\hline
\begin{tabular}[c]{@{}c@{}}$P_{\mathrm{base}}$\\(kW)\end{tabular}
& \begin{tabular}[c]{@{}c@{}}Delayed\\jobs (\%)\end{tabular}
& \begin{tabular}[c]{@{}c@{}}Mean\\delay (h)\end{tabular}
& \begin{tabular}[c]{@{}c@{}}P90\\delay (h)\end{tabular}
& \begin{tabular}[c]{@{}c@{}}Maximum\\delay (h)\end{tabular}
& \begin{tabular}[c]{@{}c@{}}Missed\\jobs\end{tabular} \\
\hline
25 & 3.118 & 1.111 & 2 & 2 & 0 \\
35 & 3.174 & 1.011 & 1 & 2 & 0 \\
45 & 3.757 & 1.000 & 1 & 1 & 0 \\
55 & 3.473 & 1.018 & 1 & 2 & 0 \\
\hline
\end{tabular}
\end{table}

\begin{table}[ptb]
\centering
\caption{PUE under different base IT powers.}
\label{tab:base_power_sensitivity_pue}
\begin{tabular}{c c}
\hline
$P_{\mathrm{base}}$ (kW) & PUE \\
\hline
25 & 1.0364 \\
35 & 1.0344 \\
45 & 1.0356 \\
55 & 1.0331 \\
\hline
\end{tabular}
\end{table}

\subsection{Impacts of Battery Capacity}
\begin{figure}[ptb]
    \centering
    \includegraphics[width=0.9\linewidth]{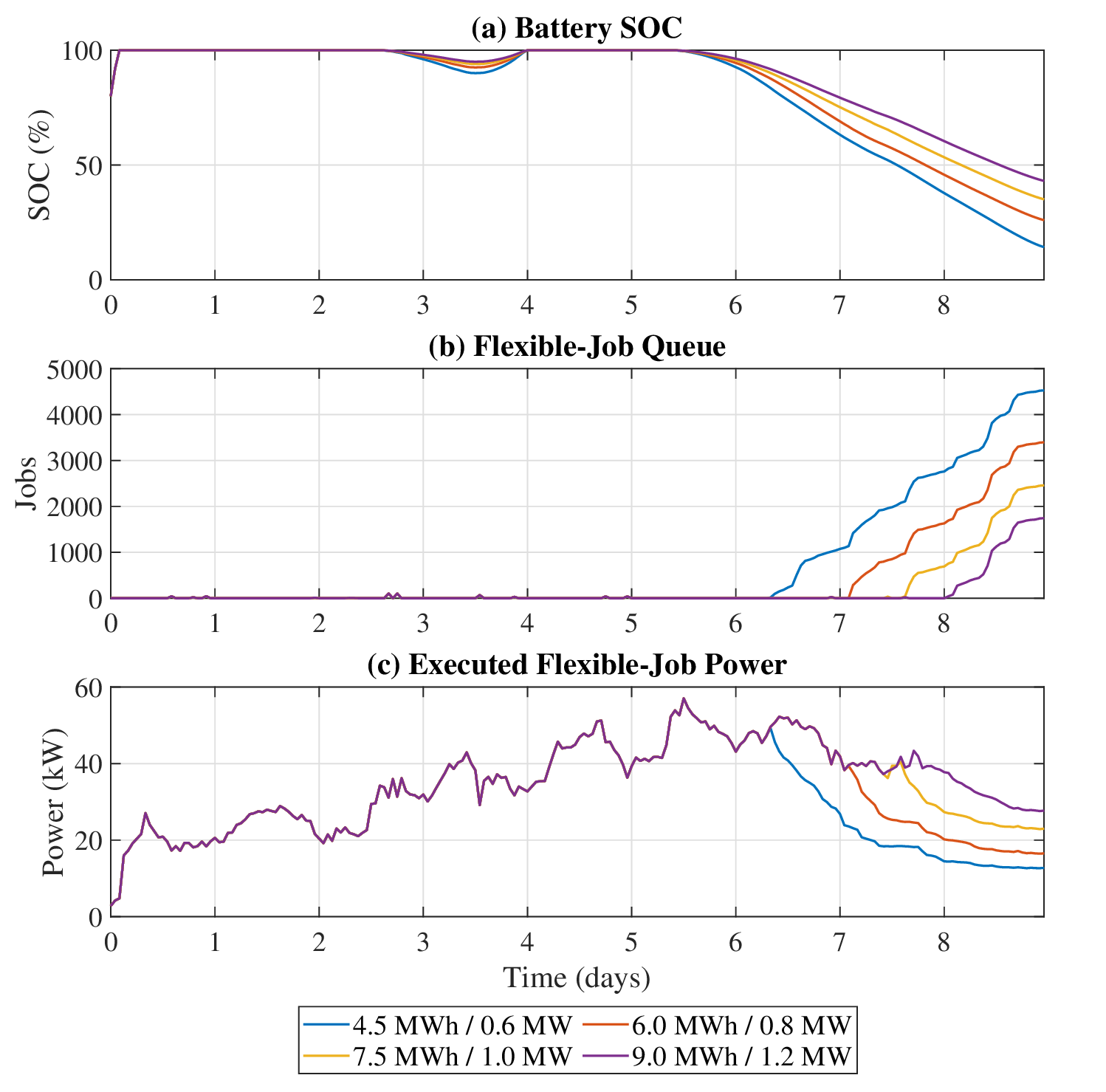}
    \caption{Closed-loop trajectories under different battery energy--power pairs.}
    \label{fig:battery_pair_closed_loop}
\end{figure}

\begin{figure}[ptb]
    \centering
    \includegraphics[width=0.9\linewidth]{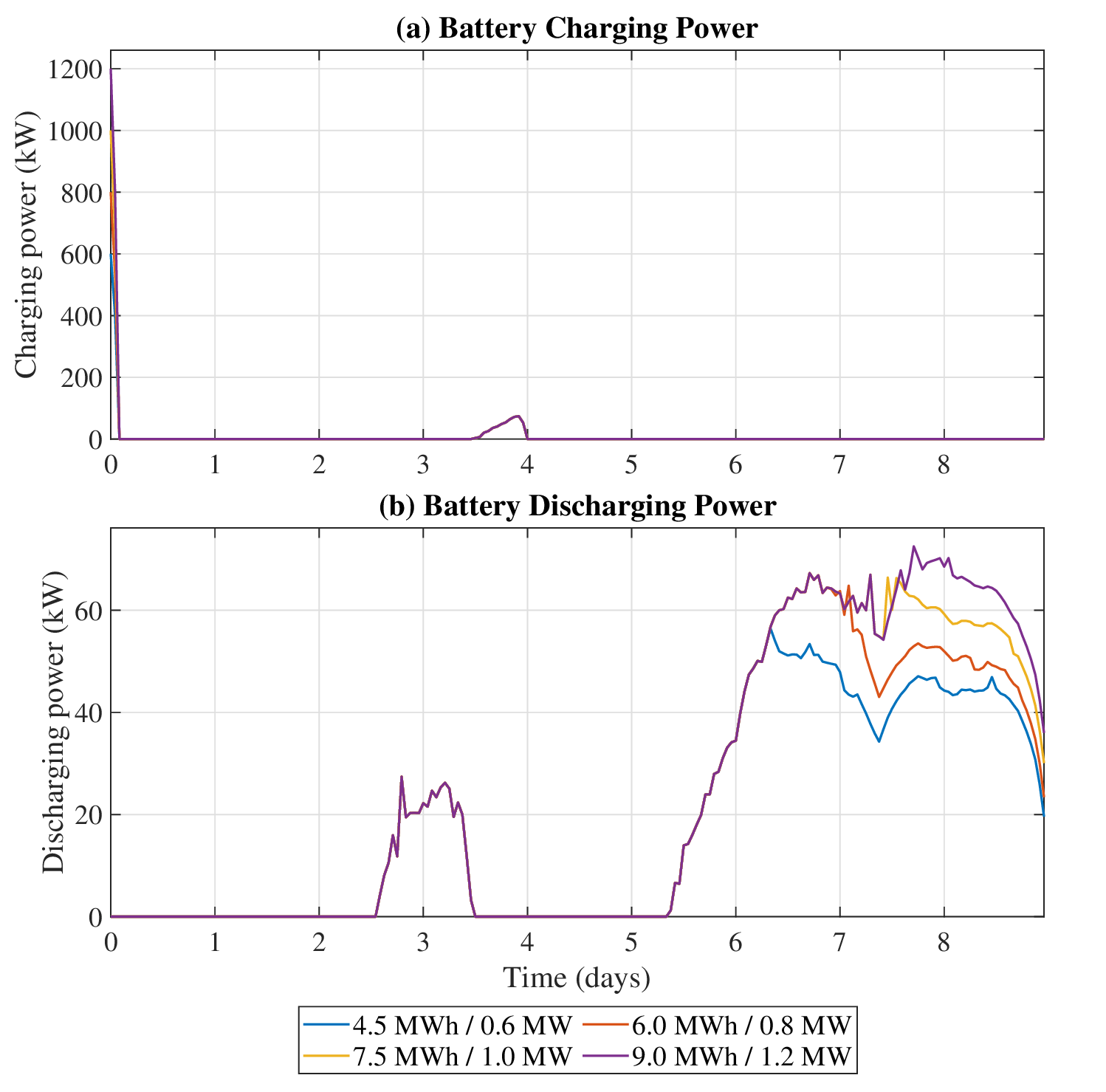}
    \caption{Battery power under different battery energy--power pairs.}
    \label{fig:battery_pair_power}
\end{figure}

The fifth study evaluates the influence of the battery capacity (both storage capacity and charging/discharging capacity) while keeping the workload, environmental conditions, WEC configuration, thermal model, and controller parameters unchanged, and the week beginning on July 17 is evaluated in all cases. Four battery configurations with the same energy-to-power ratio are considered. Figure~\ref{fig:battery_pair_closed_loop} presents the corresponding battery SOC, flexible-job queue, and executed flexible-job power trajectories, while Figure~\ref{fig:battery_pair_power} shows the charging and discharging power. As shown in Figure~\ref{fig:battery_pair_closed_loop}, larger battery capacities maintain a higher SOC near the end of the simulation and smaller accumulation of flexible jobs in the queue. The executed flexible-job power is similar among all cases during the earlier period, but larger battery capacities support higher flexible-job power after the SOC trajectories begin to separate. Figure~\ref{fig:battery_pair_power} further shows that the charging profiles are similar, whereas larger battery capacities provide higher discharging power during the later workload period.

The job-level QoS results are summarized in Table~\ref{tab:battery_size_sensitivity_qos}. Increasing the battery capacity from 4.5/0.6 MWh/MW to 6.0/0.8 MWh/MW reduces the delayed-job percentage and eliminates missed jobs. Further increases in battery size do not improve the reported delay metrics, since the 6.0/0.8, 7.5/1.0, and 9.0/1.2 MWh/MW cases produce the same job-level QoS results. Therefore, 6.0/0.8 MWh/MW is selected as the smallest evaluated battery pair that avoids missed jobs while achieving the same reported QoS as the larger configurations.
\begin{table}[ptb]
\centering
\caption{Flexible-job QoS under different battery size pairs.}
\label{tab:battery_size_sensitivity_qos}
\small
\setlength{\tabcolsep}{3pt}
\begin{tabular}{c c c c c c}
\hline
\begin{tabular}[c]{@{}c@{}}Battery\\(MWh / MW)\end{tabular}
& \begin{tabular}[c]{@{}c@{}}Delayed\\jobs (\%)\end{tabular}
& \begin{tabular}[c]{@{}c@{}}Mean\\delay (h)\end{tabular}
& \begin{tabular}[c]{@{}c@{}}P90\\delay (h)\end{tabular}
& \begin{tabular}[c]{@{}c@{}}Maximum\\delay (h)\end{tabular}
& \begin{tabular}[c]{@{}c@{}}Missed\\jobs\end{tabular} \\
\hline
4.5 / 0.6 & 10.806 & 1.000 & 1 & 1 & 1041 \\
6.0 / 0.8 & 3.757  & 1.000 & 1 & 1 & 0 \\
7.5 / 1.0 & 3.757  & 1.000 & 1 & 1 & 0 \\
9.0 / 1.2 & 3.757  & 1.000 & 1 & 1 & 0 \\
\hline
\end{tabular}
\end{table}

\subsection{Impacts of WEC Configuration}

\begin{figure}[ptb]
    \centering
    \includegraphics[width=0.9\linewidth]{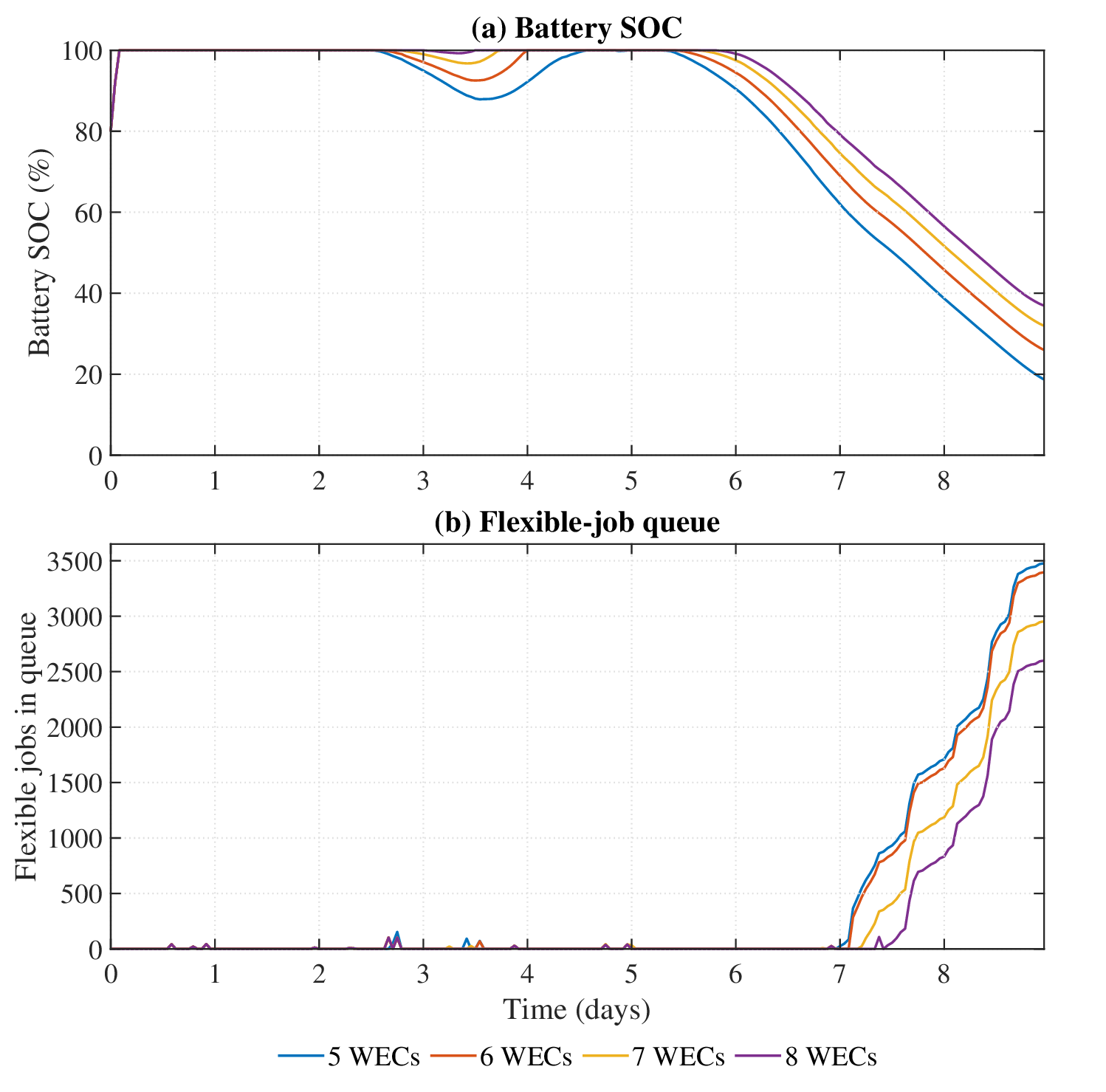}
    \caption{Battery-SOC and flexible-job queue trajectories under different numbers of WEC units.}
    \label{fig:wec_count_soc_queue}
\end{figure}

\begin{figure}[ptb]
    \centering
    \includegraphics[width=0.9\linewidth]{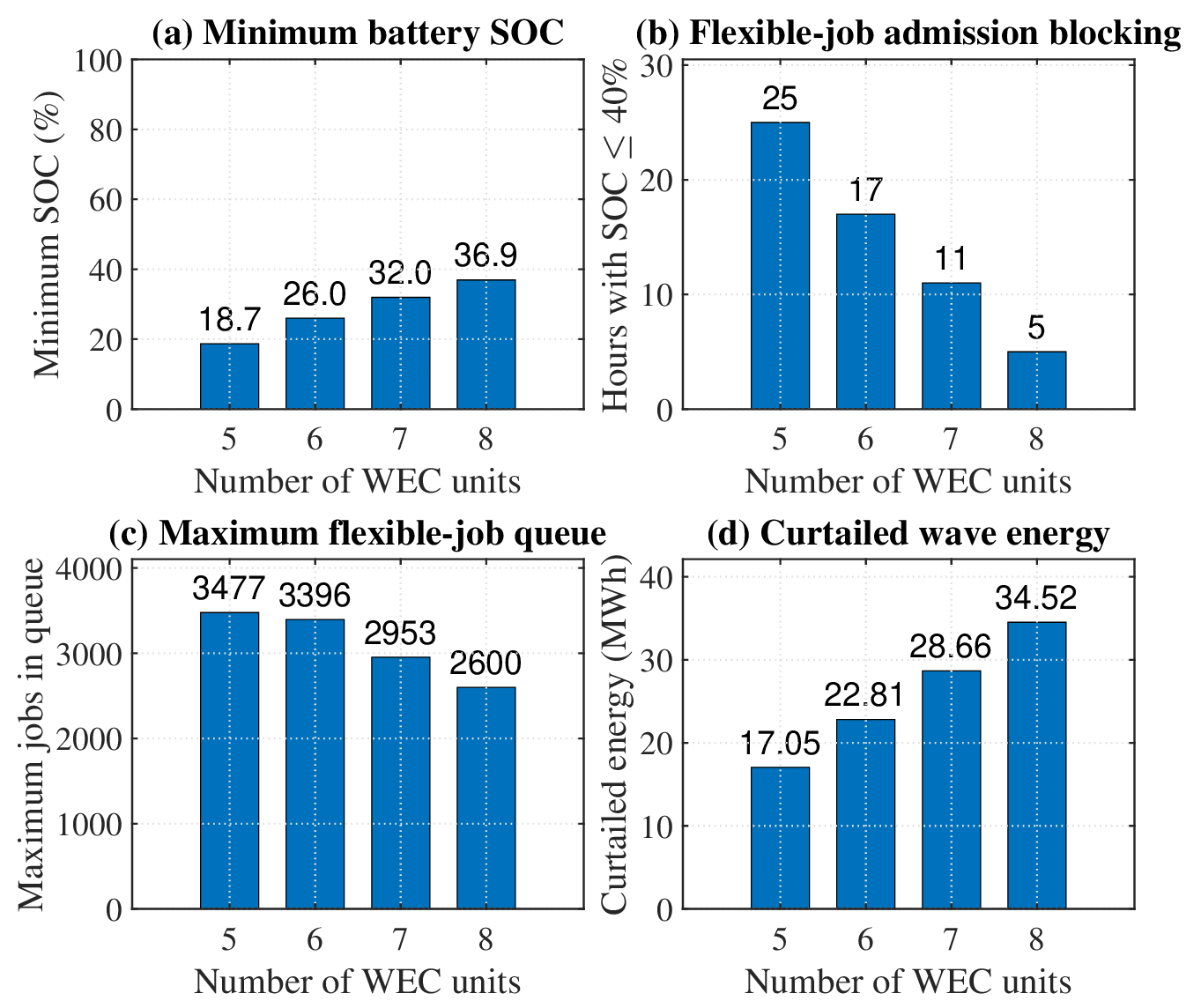}
    \caption{Battery, workload-scheduling, and wave-energy-curtailment results under different numbers of WEC units.}
    \label{fig:wec_count_summary}
\end{figure}
\begin{table}[ptb]
\centering
\caption{Flexible-job QoS under different numbers of WEC units.}
\label{tab:wec_count_sensitivity_qos}
\small
\setlength{\tabcolsep}{3pt}
\begin{tabular}{c c c c c c}
\hline
\begin{tabular}[c]{@{}c@{}}Number\\of WECs\end{tabular}
& \begin{tabular}[c]{@{}c@{}}Delayed\\jobs (\%)\end{tabular}
& \begin{tabular}[c]{@{}c@{}}Mean\\delay (h)\end{tabular}
& \begin{tabular}[c]{@{}c@{}}P90\\delay (h)\end{tabular}
& \begin{tabular}[c]{@{}c@{}}Maximum\\delay (h)\end{tabular}
& \begin{tabular}[c]{@{}c@{}}Missed\\jobs\end{tabular} \\
\hline
5 & 4.028 & 1.093 & 1 & 2 & 0 \\
6 & 3.757 & 1.000 & 1 & 1 & 0 \\
7 & 3.875 & 1.014 & 1 & 2 & 0 \\
8 & 3.257 & 1.000 & 1 & 1 & 0 \\
\hline
\end{tabular}
\end{table}
The sixth study evaluates the influence of the number of WEC units while keeping the workload, environmental conditions, battery configuration, thermal model, and controller parameters unchanged, and the week beginning on July 17 is evaluated in all cases. Four WEC configurations are considered. Figure~\ref{fig:wec_count_soc_queue} presents the corresponding battery SOC and flexible-job queue trajectories, while Figure~\ref{fig:wec_count_summary} summarizes the minimum SOC, flexible-job admission blocking, maximum queue size, and curtailed wave energy. The curtailed wave energy refers to the available wave energy that cannot be directly consumed by the IT and cooling loads or stored in the battery because of the battery SOC and charging-power limits. As shown in Figure~\ref{fig:wec_count_soc_queue}, increasing the number of WEC units maintains a higher battery SOC during the later workload period and reduces the accumulation of flexible jobs in the queue. Figure~\ref{fig:wec_count_summary} further shows that the minimum SOC increases, the time below the new-flex admission threshold decreases, and the maximum queue size is reduced as more WEC units are installed. However, the curtailed wave energy also increases, indicating that part of the additional renewable energy cannot be fully utilized.

The job-level QoS results are summarized in Table~\ref{tab:wec_count_sensitivity_qos}. The delayed-job percentage varies only slightly among the evaluated cases and is not strictly monotonic for the intermediate WEC configurations, while the largest WEC configuration achieves the lowest delayed-job percentage. The mean and P90 delays remain similar, and no missed jobs occur in any case. Therefore, increasing the number of WEC units mainly improves battery availability and limits queue accumulation, whereas the improvement in job-level delay is relatively small and is accompanied by increased wave-energy curtailment.

\subsection{Monte Carlo Simulation and Sensitivity Analysis}

\begin{figure}[ptb]
    \centering
    \includegraphics[width=0.9\linewidth]{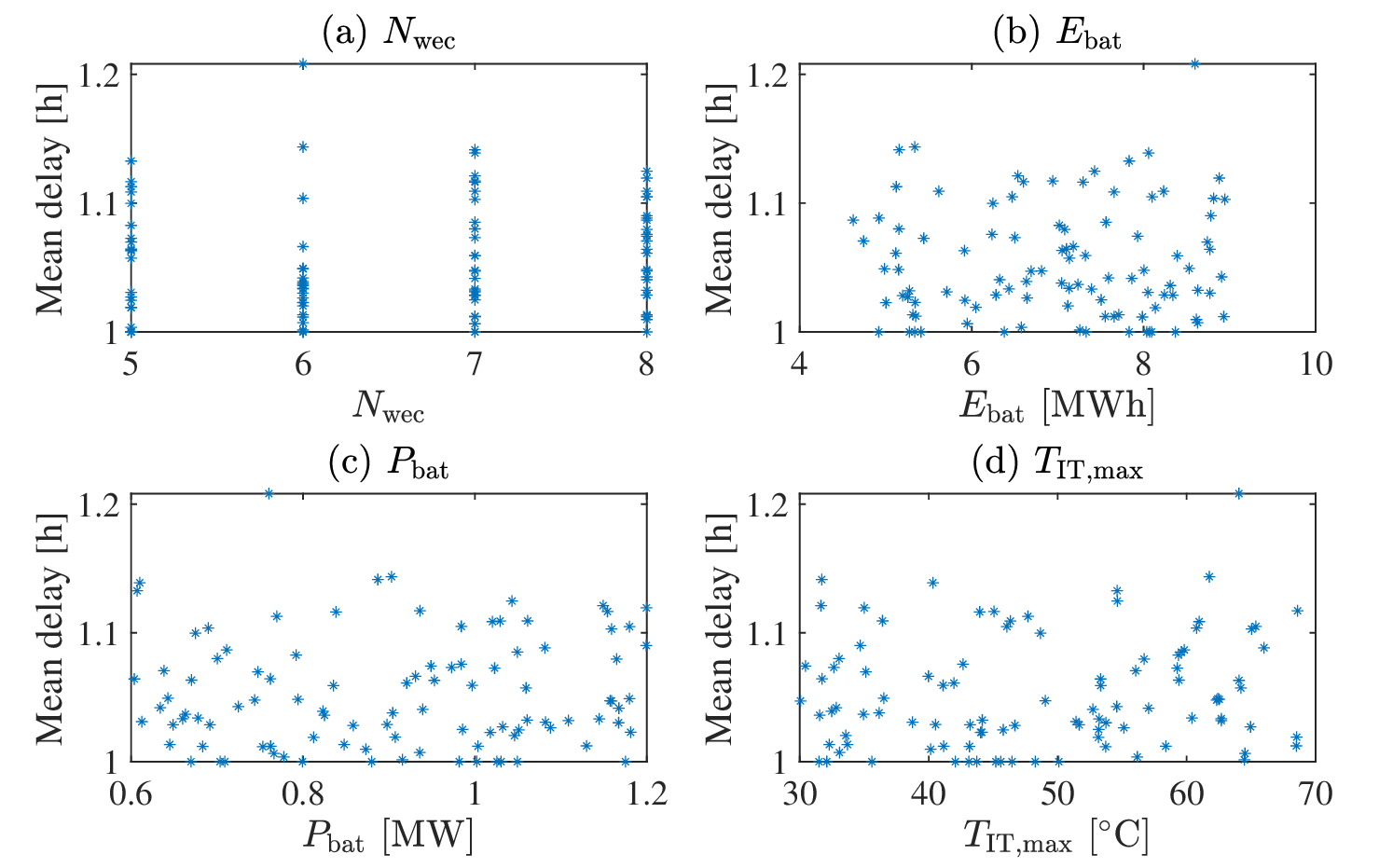}
    \caption{Monte Carlo simulation results for mean flexible-job delay under randomly sampled system parameters.}
    \label{fig:monte_carlo_mean_delay}
\end{figure}

\begin{figure}[ptb]
    \centering
    \includegraphics[width=0.9\linewidth]{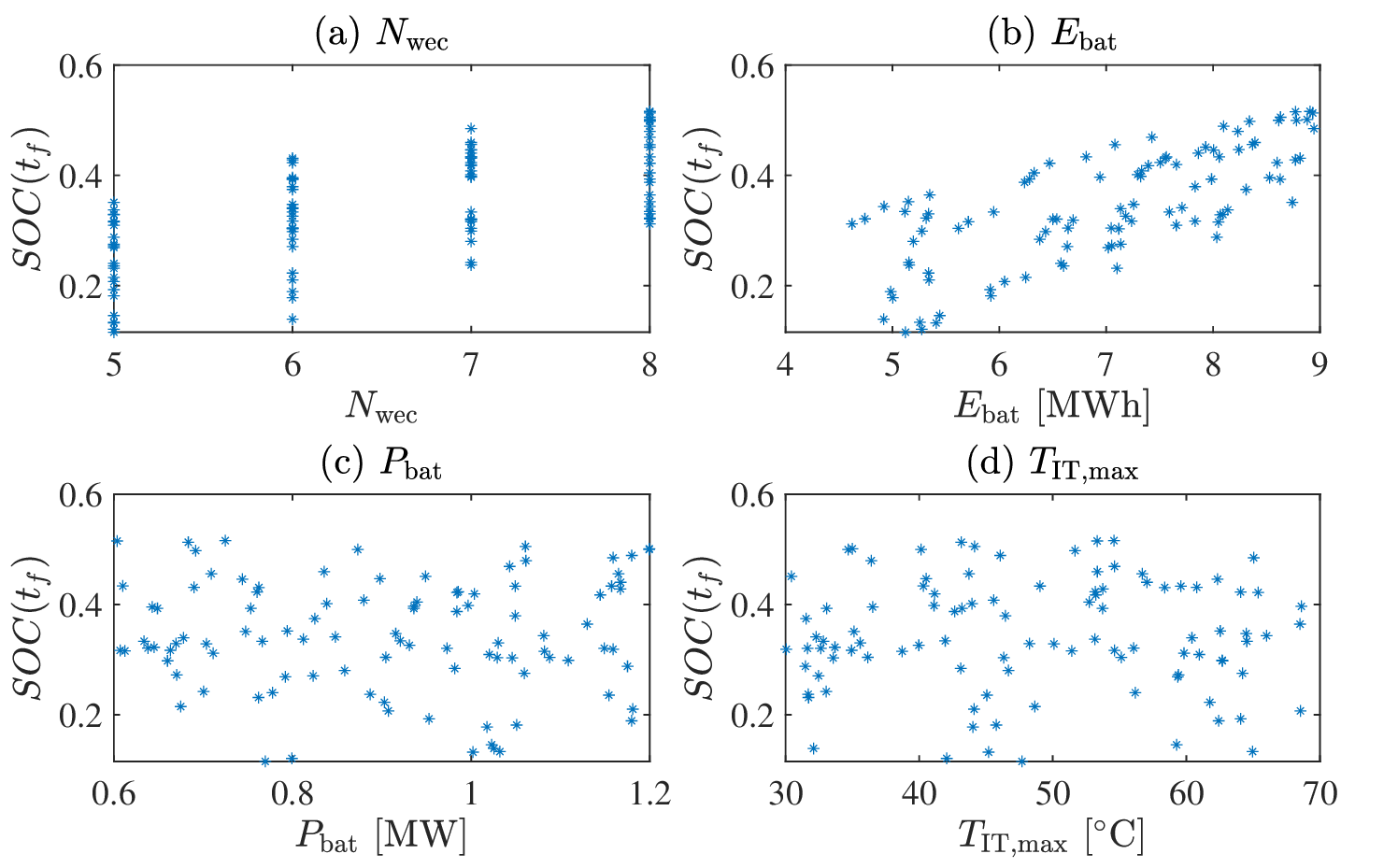}
    \caption{Monte Carlo simulation results for terminal battery SOC under randomly sampled system parameters.}
    \label{fig:monte_carlo_terminal_soc}
\end{figure}

A Monte Carlo study is conducted to evaluate the dependence of the closed-loop performance on the renewable-generation, battery-sizing, and thermal-limit parameters. A total of 100 simulations are performed. For each simulation, the number of WEC units $N_{\mathrm{wec}}$ is randomly selected from $\{5,6,7,8\}$, the battery energy capacity $E_{\mathrm{bat}}$ is randomly sampled from $4.5$--$9~\mathrm{MWh}$, the battery power capacity $P_{\mathrm{bat}}$ is randomly sampled from $0.6$--$1.2~\mathrm{MW}$, and the maximum allowable IT temperature $T_{\mathrm{IT,max}}$ is randomly sampled from $30$--$70^{\circ}\mathrm{C}$. All remaining workload, environmental, thermal-model, and controller settings are kept unchanged. For each case, the mean delay of the delayed flexible jobs and the terminal battery SOC are reported.

To quantify the statistical dependence between each sampled parameter and the two performance metrics, the squared distance correlation $dCor^2$ is calculated following~\cite{chen2023battery,szekely2007measuring,szekely2009brownian}. For two random variables $X$ and $Y$, it is defined as
\begin{equation}
dCor^2(X,Y)=\frac{dCov^2(X,Y)}{\sqrt{dVar^2(X)dVar^2(Y)}},
\end{equation}
where
\begin{equation}
dVar^2(X)=dCov^2(X,X).
\end{equation}
The detailed sample computation follows~\cite{chen2023battery,szekely2007measuring,szekely2009brownian}.

\begin{table}[ptb]
\centering
\caption{Squared distance-correlation results for the Monte Carlo simulations.}
\label{tab:monte_carlo_dcor}
\small
\begin{tabular}{c c c}
\hline
Parameter & Mean delay & Terminal SOC \\
\hline
$N_{\mathrm{wec}}$    & 0.039 & 0.409 \\
$E_{\mathrm{bat}}$    & 0.010 & 0.486 \\
$P_{\mathrm{bat}}$    & 0.017 & 0.026 \\
$T_{\mathrm{IT,max}}$ & 0.028 & 0.023 \\
\hline
\end{tabular}
\end{table}

The Monte Carlo results are shown in Figs.~\ref{fig:monte_carlo_mean_delay} and \ref{fig:monte_carlo_terminal_soc}, while the corresponding squared distance-correlation values are summarized in Table~\ref{tab:monte_carlo_dcor}. As shown in Fig.~\ref{fig:monte_carlo_mean_delay}, the mean delay remains concentrated near $1~\mathrm{h}$ over the sampled parameter
ranges, and no clear trend is observed with respect to any of the four parameters. This is consistent with Table~\ref{tab:monte_carlo_dcor}, where the $dCor^2$ values for $N_{\mathrm{wec}}$, $E_{\mathrm{bat}}$, $P_{\mathrm{bat}}$, and $T_{\mathrm{IT,max}}$ are 0.039, 0.010, 0.017, and 0.028, respectively. These results indicate that the mean delay of delayed flexible jobs is relatively insensitive to the sampled system parameters over the investigated ranges. In contrast, Fig.~\ref{fig:monte_carlo_terminal_soc} shows a clearer dependence of the terminal battery SOC on $E_{\mathrm{bat}}$ and $N_{\mathrm{wec}}$. The terminal SOC generally increases with both battery energy capacity and the number of WEC units. Consistently,
Table~\ref{tab:monte_carlo_dcor} gives $dCor^2$ values of 0.486 and 0.409 for $E_{\mathrm{bat}}$ and $N_{\mathrm{wec}}$, respectively, whereas the corresponding values for $P_{\mathrm{bat}}$ and $T_{\mathrm{IT,max}}$ are only 0.026 and 0.023. This indicates that the terminal battery SOC is more strongly associated with battery energy capacity and wave-generation capacity, while its dependence on
battery power capacity and the IT temperature limit is much weaker.

\section{Conclusion}\label{sec-conclusion}

This paper developed an integrated NMPC framework for coordinating cooling operation, flexible-workload scheduling, wave-power utilization, and battery management in a wave-powered subsea data center. A workload model constructed from job-level CPU, memory, and GPU data, together with a three-node thermal model, was incorporated into the closed-loop control framework. Simulation results demonstrate that the proposed controller maintains the IT temperature within its prescribed limit while adapting cooling operation and flexible-workload execution according to wave-power availability and battery SOC. The parametric studies show that workload and environmental variations can significantly affect battery utilization, queue accumulation, and flexible-job service performance. Increasing the base IT power primarily reduces the battery-energy margin and raises the IT operating temperature, while larger battery configurations improve workload-service reliability. Increasing the number of WEC units improves battery availability and reduces flexible-workload queue accumulation, but also increases wave-energy curtailment. The Monte Carlo simulation and sensitivity analyses further show that the mean delay of delayed flexible jobs is relatively insensitive to the investigated system parameters, whereas terminal battery SOC is primarily influenced by battery energy capacity and wave-generation capacity. Overall, the proposed framework provides an integrated approach for evaluating and coordinating the coupled thermal, workload, renewable-generation, and energy-storage behavior of wave-powered subsea data centers. {Future work will focus on full-year simulations to determine appropriate battery sizing under seasonal variations in workload, wave-power generation, and seawater temperature. Further developments will include higher-fidelity thermal and cooling models, more flexible workload-scheduling strategies, and joint optimization of energy-storage sizing and system operation.}



\balance
\bibliographystyle{IEEEtran}
\bibliography{ref}

\end{document}